\documentclass{jfm}

\usepackage{graphicx} 

\usepackage{amsmath}
\usepackage{amssymb}
\usepackage{bm}
\usepackage{bbm}
\usepackage{comment}
\usepackage{tikz}
\usetikzlibrary{backgrounds, intersections}

\newcommand{\diff}{\mathrm{d}}

\title{Stability of trapped fluid clusters in two-phase porous media flow}

\author{Mathias Klahn\aff{1}, Gaute Linga\aff{2,3}, Tanguy Le Borgne\aff{2,4}, Joachim Mathiesen\aff{1,2}}

\affiliation{\aff{1}Niels Bohr Institute, University of Copenhagen, Copenhagen, Denmark, 
\aff{2}The Njord Centre, Departments of Geosciences and Physics, University of Oslo, Oslo, Norway, 
\aff{3}PoreLab, Department of Physics, Norwegian University of Science and Technology, Trondheim, Norway,
\aff{4}Géosciences Rennes, UMR 6118 CNRS University of Rennes, Rennes, France}

\begin{document}

\maketitle

\begin{abstract}
A key challenge in multiphase flow through porous media is to understand and predict the conditions under which trapped fluid clusters become mobilized. Here, we investigate the stability of such clusters in two-phase flow and present a simple, quasistatic model that accurately determines the critical Bond number (that is, the critical ratio between the average pressure gradient of the flow and the surface tension) for the onset of cluster mobilization. The model is derived by combining elementary geometrical considerations with mass conservation and a mechanical equilibrium condition, resulting in a system of coupled differential equations. Our derivation sheds new light on the mechanisms that govern cluster stability. In addition, since the number of equations equals the number of cluster openings, our model is significantly faster than direct numerical simulations of the same problem and enables efficient exploration of the system's parameter space. Using this approach, we highlight a discrepancy with the prediction of current mean-field theories, which predict that the largest stable cluster size scales in proportion to $r/\text{Bo}^\alpha$, where $r$ is a typical pore size, $\text{Bo}$ is the Bond number and $\alpha$ is a fixed exponent. We discuss the mechanisms that explain the breakdown of the mean field theories, and we show that a scaling law of this form can only exist if $\alpha$ is allowed to depend on a broad set of flow and geometric parameters.
\end{abstract}

\section{Introduction}\label{sec:introduction}
The motion of two immiscible fluids through a porous media is known to fall into different regimes, depending on the flow's ability to mobilize trapped clusters \citep{AvraamPayatakes1995}. As such, the stability of trapped clusters is important for our basic understanding of two-phase porous-media flow in general and hence also for a long list of industrial applications. While enhanced oil recovery has traditionally dominated this list, many other important applications now also motivate the study of this problem; it is, for example, relevant to water or contaminant infiltration in soil, solute transport in unsaturated systems and soil remediation, and for discussions of these subjects we refer the reader to the papers of \cite{KirchnerEtAl2000}, \cite{ZhanNg2004}, \cite{AlamootiEtAl2023} and \cite{RahbehMohtar2007}. Naturally, the problem's importance has led several authors to suggest solutions, and, being based on mean-field arguments, these solutions not only improve our understanding, they also provide easy-to-use formulas for predicting whether a trapped cluster will be stable under given flow conditions. These formulas are, however, not always accurate, and since they are not derived from first principles, their range of validity is not known. In addition, as we will argue in the following, the existing mean-field arguments neglect several important features that become apparent when studying cluster stability on the pore scale.

A first step towards a characterization of cluster stability was taken by \cite{Taber1969}, who experimentally showed that the fraction of trapped clusters that can be mobilized by a given flow is a function of the average pressure gradient divided by the surface tension between the two fluids. A few years later, \cite{MelroseBrandner1974} and \cite{LarsonEtAl1977} provided a mean-field argument suggesting that a trapped cluster cannot be stable if its size exceeds a numerical factor determined by the porous geometry times  $r/\text{Bo}$, where $r$ corresponds to a typical pore size and $\text{Bo}$ is the Bond number measuring the strength of the average pressure gradient of the flow relative to the surface tension; exact definitions of these quantities are given in Section \ref{sec:physicalSystem} below. Throughout this paper, we shall refer to this argument as MFA1, and we note that it was successfully applied by \cite{TallakstadEtAl2009} to explain the results of their measurements of cluster sizes in statistically steady flows. The argument first considers the pressure difference across the cluster and assumes that this quantity must equal the average pressure gradient times the size of the cluster. Secondly, based on the Young-Laplace equation, MFA1 assumes that the pressure difference that can be sustained by the trapped cluster is on the order of the surface tension divided by a typical pore size. Finally, equating the two pressure differences then gives the $r/\text{Bo}$-dependence of the largest stable cluster size. While this argument is both simple and elegant, it however also neglects several important physical features of the problem. First of all, the assumption that a trapped cluster becomes unstable when the difference in pressure outside its most upstream (bottom) and downstream (top) openings exceeds the pressure difference that can be sustained by the average pore opening is questionable. The relative strength of two different cluster openings is determined by a competition between their widths as well as the inside and outside pressures that the openings are exposed to. As such, it is often not the opening located the furthest upstream that breaks first, and therefore the relevant pressure difference is not necessarily the one between the bottom and the top. Moreover, the assumption disregards the fact that a large cluster is more likely to contain a wide (and therefore weak) opening than a small cluster, and one would therefore expect this effect to contribute to the fact that large clusters in general are more easily mobilized than small clusters.

The effect was described by \cite{MathiesenEtAl2023}, who put forth an argument, to which we shall refer to as MFA2 in the remainder of this paper. This argument predicts that the largest stable cluster size should scale as $r/\text{Bo}^{1/2}$ and as such, it predicts more clusters to be unstable than MFA1, since mobilization takes place for Bond numbers smaller than one. MFA2 differs from MFA1 by assuming that a cluster becomes unstable when its widest—and therefore weakest—opening can no longer sustain the pressure difference across the cluster. However, the maximum pressure difference that a random opening can withstand is assumed to follow an exponential distribution, and the $r/\text{Bo}^{1/2}$-scaling turns out to be a result of the specific form of this distribution. Unfortunately, presently no theoretical argument supports this specific form, and by assuming a different distribution, one may use the argument to predict that the largest stable cluster size should scale as $r/\text{Bo}^\alpha$ for any desired value of $\alpha$. In addition, MFA2 suffers from other limitations, as it assumes that cluster openings can be considered independently and that the maximum pressure difference that an opening can sustain follows a simple universal distribution.

So far we have described shortcomings that are specific to MFA1 and MFA2. A fundamental concern, which applies to any mean-field approach, is that the outside pressure field is necessarily approximated through the average pressure gradient of the flow. In other words, the random fluctuations of the pressure that appear on the pore scale are neglected. Since these fluctuations are mainly induced by the stochastic nature of the porous geometry, their typical size is roughly independent of the cluster size, implying that the fluctuation-to-mean ratio diverges in the limit where the cluster becomes small. As such, the mean-field approximation for the outside pressure can only be accurate for large clusters.

In summary, the stability of trapped clusters remains an open question; different scaling relations have been proposed, and their derivations are based on physical assumptions that do not universally hold and which do not take some key pore-scale mechanisms into account. The reason for this is, of course, the complicated, nonlinear nature of two-phase porous-media flow, which renders exact theoretical treatments very challenging and at the same time makes detailed experimental and numerical investigations both difficult and extremely time-consuming. However, for some specific applications, there exists analytical models succeeding in only taking the most relevant physical phenomena into account, allowing them to offer both accurate quantitative predictions as well as new physical insights. A prime example of this is the very simple stepwise fluid propagation model proposed by \cite{CieplakRobbins1988} and \cite{CieplakRobbins1990}. Relying only on elementary geometric considerations, these authors showed that a cluster interface can become unstable in three different ways, and by combining this with ``a set of plausible rules'' for the formation of new interfaces following instability, as \cite{SinghEtAl2019} put it in their review, they managed to explain the role played by the contact angle during invasion processes. The validity of Cieplak \& Robbins' classification of the modes of instability has since been confirmed experimentally by \cite{JungEtAl2016}, and the recent work presented by, for example, \cite{HuEtAl2019}, \cite{PrimkulovEtAl2021} and \cite{WangEtAl2023}, is a testament to the fact that simple geometrical ideas continue to play an important role in two-phase porous media flow.

In this work, we consider a trapped cluster in a two-dimensional porous medium and develop a new, quasistatic model for studying its stability. While this model can also be derived for a three-dimensional system, we have chosen the two-dimensional setting because it allows us to highlight the model's physical basis, which is easily overlooked in three-dimensional porous geometries. Since MFA1 and MFA2 apply equally well to two and three dimensions, the decision to consider a two-dimensional system is not a barrier to testing the validity of these arguments. Moreover, two-dimensional porous media are broadly used and represent an important reference in experiments, simulations and theory. The model consists of a set of coupled differential equations predicting how the geometry of the cluster changes with various parameters up to the point of breaking, and we derive it by making geometrical considerations similar to those of \cite{CieplakRobbins1988} and by requiring mass conservation and that the cluster is in mechanical equilibrium at all times. As such, the derivation sheds new light on the fundamental understanding of cluster stability. Moreover, as the equations are much faster to solve than the full hydrodynamic problem, they provide a vehicle for exploring a much larger subset of the system's parameter space than what is currently possible with direct numerical simulation. We use our model to show that the scaling laws for the largest stable cluster size predicted by the mean-field arguments MFA1 and MFA2 do not hold in general. Furthermore, we question the usefulness of expressing the largest stable cluster size on the form $r/\text{Bo}^\alpha$, where $\alpha$ is some exponent; for as we will show, if one insists on this form, $\alpha$ is necessarily a complicated function of several different variables.

The remainder of this paper is laid out as follows: in Section \ref{sec:physicalSystem}, we specify the porous-media system under consideration. In Section \ref{sec:simplifiedModel}, we develop the quasistatic model for predicting the evolution of the cluster geometry with respect to the Bond number and explain how to extract the point of breaking from this model. The model is validated against direct numerical simulations in Section \ref{sec:validation}, and subsequently used to compute the probability that a cluster of a given size is stable as a function of the Bond number in Section \ref{sec:stabilityProb}. Conclusions are drawn in Section \ref{sec:conclusions}.

\section{Physical system}\label{sec:physicalSystem}
We consider the stability of a single, trapped cluster of wetting (w) fluid surrounded by non-wetting (nw) fluid in a two-dimensional gravity-driven porous-media flow. We assume both fluids to be incompressible and immiscible, and we denote their dynamic viscosities by $\mu_\mathrm{w}$ and $\mu_\mathrm{nw}$, and their mass densitites by $\rho_\mathrm{w}$ and $\rho_\mathrm{nw}$. The fluids are both subject to the gravitational acceleration $\bm{g}$, which we take to point in the positive $y$-direction such that $\bm{g} = (0, g)$. Moreover, we let $\gamma$ and $\theta_0$ represent the surface tension between the fluids and the contact angle, respectively.

\begin{figure}
    \centering
    \input{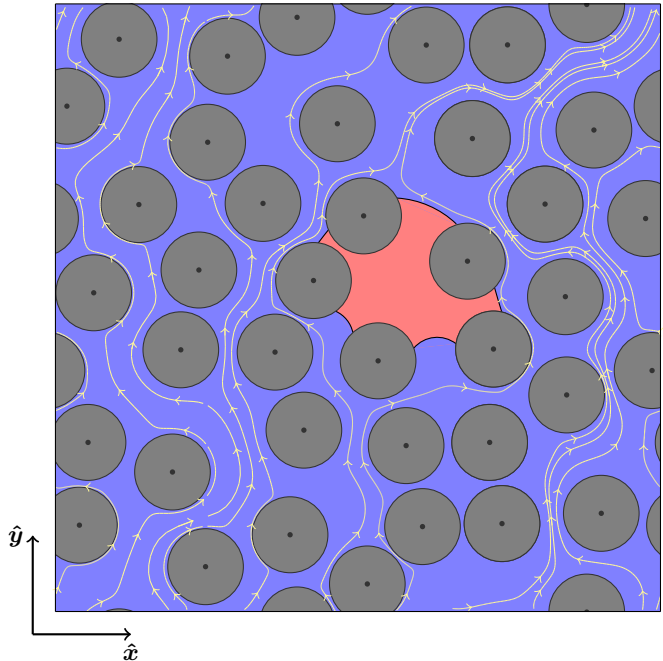}
    \caption{A cluster of wetting fluid (red) surrounded by non-wetting fluid (blue) trapped between the cylindrical obstacles of the porous medium. The Bond number is smaller than the critical value such that the cluster lies still. In contrast, the non-wetting fluid flows according to a time-independent velocity field. The streamlines of the flow are illustrated by the yellow curves, which are actual streamlines taken from numerical simulations.}
    \label{fig:physicalSystem2D}
\end{figure}

For the porous media, we consider idealized geometries consisting of square domains filled with non-overlapping, cylindrical obstacles of radius $r$. Realizations of the geometries are generated using the random sequential adsorption procedure (see, for example, \citet{Feder1980}) which is commonly used to construct random isotropic porous media with a single length scale. In this procedure, new obstacles are successively placed at random positions drawn from a uniform distribution over the domain such that the distance between any two obstacle centers is at least $d_{\text{min}} + 2r$; the process terminates when this is no longer possible. Throughout this work we use the value $d_\text{min} = r/10$ for the minimum distance, and we note that this implies that the porosity of the medium is close to $50 \%$. Figure \ref{fig:physicalSystem2D} shows a sketch of the system, including the trapped cluster and the streamlines of the flow. 

We fix the properties of the fluids and obstacles such that the flow around the trapped cluster is creeping and consider the stability of the cluster in terms of the Bond number
\begin{equation}\label{eq:BondNumber}
    \text{Bo} \equiv \frac{\rho_\text{w} g r^2}{\gamma}.
\end{equation}
To be precise, we study the stability of trapped clusters in the following way: first, pick a trapped cluster and choose the Bond number small enough that it is stable. Secondly, start increasing the Bond number very slowly, that is, over a time scale much longer than the time it takes the flow around the cluster to readjust to the new conditions. During this process, the cluster remains in mechanical equilibrium the entire time and the geometry of the cluster changes until an opening of the cluster finally becomes unstable and is pushed in. We call the Bond number where this happens the critical Bond number and denote it by $\text{Bo}_\text{crit}$. In this connection, we note that one could, of course, define the critical Bond number in various ways, for example as the smallest value of $\text{Bo}$ where the cluster is fully mobilized, i.e. completely pushed away from the region of space that it occupies in the limit $\text{Bo} \rightarrow 0$. This definition, however, easily becomes relatively involved, and since it produces critical Bond numbers which are typically quite close to the ones corresponding to the first unstable opening, we have chosen to use the definition based on this criterion in this work.

\section{A quasistatic model for predicting the critical Bond number}\label{sec:simplifiedModel}
\begin{figure}
    \centering
    \input{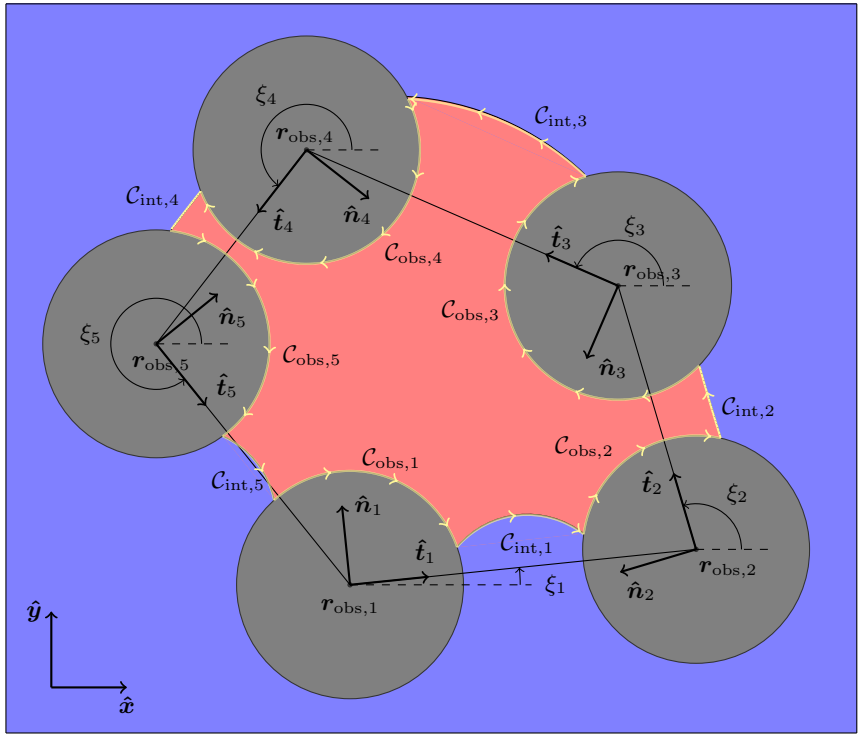}
    \caption{A cluster trapped between obstacles. The obstacles are labeled in counterclockwise direction, and the center of the $n$th obstacle is at $\bm{r}_{\text{obs},n}$. Each pair of obstacle neighbours constitutes a cluster opening and is equipped with a local coordinate system, $\{ \bm{\hat{t}}_n, \bm{\hat{n}}_n\}$. The angle between $\bm{\hat{t}}_n$ and the $x$-axis is denoted by $\xi_n$. The curve $\mathcal{C}$ follows the circumference of the cluster and is oriented counterclockwise; for computational purposes this curve is decomposed into segments along the fluid interfaces, $\mathcal{C}_{\text{int}, n}$, and segments along the obstacles, $\mathcal{C}_{\text{obs},n}$.}   
    \label{fig:clusterParameterization}
\end{figure}

\begin{figure}
    \centering
    \begin{tikzpicture}
    \def\r{2}
    \def\d{6}
    \def\omegaDeg{25}
    \def\omegaRad{5*pi/36}

    \def\xL{0.0}
    \def\yL{0.0}
    
    \def\xR{\d+2.0*\r}
    \def\yR{0.0}

    \def\R{3.89108} 
    \def\xC{\r + 0.5*\d}
    \def\yC{-1.38660} 

    \fill[red!50] (\xL-\r,\yL) rectangle (\d+3*\r, 1.7*\r);
    \fill[blue!50] (\xL-\r,\yL) rectangle (\d+3*\r, -2.8*\r);
    \fill[blue!50] (\xC, \yC) -- ++(0:\R) arc[start angle=0, end angle=180, radius=\R] -- cycle;
    
    \fill[gray!50] (\xL, \yL) circle [radius=\r];
    \fill[gray!50] (\xR, \yR) circle [radius=\r];

    \draw (\xL, \yL) circle (\r);
    \draw (\xR, \yR) circle (\r);

    \draw[thick, ->] (\xL, \yL) -- (\xL+0.5*\r, \yL) node[anchor=north] {$\bm{\hat{t}}_n$};
    \draw[thick, ->] (\xL, \yL) -- (\xL, \yL+0.5*\r) node[anchor=east] {$\bm{ \hat{n}}_n$};
    \draw (\xL, \yL) -- (\xL + 0.8*\r, \yL);
    
    \draw (\xL, \yL) -- ({\xL + \r*cos(\omegaRad r)}, {\yL + \r*sin(\omegaRad r)});
    \draw[->] (\xL + 0.65*\r, \yL) arc[start angle=0, end angle=\omegaDeg, radius=0.65*\r]; 
    \node at (\xL + 0.8*\r, \yL+0.15*\r) {$\omega_n$};
    \node at (\xL + 0.35*\r, \yL+0.35*\r) {$r$};

    \fill (\xC, \yC) circle[radius=2pt];
    \draw (\xC, \yC) -- (\xC - \r, \yC);
    \draw[->] (\xC - 0.65*\r, \yC) arc (180:145:0.65*\r);
    \node at (\xC - 0.8*\r, \yC + 0.2*\r) {$\Omega_n$};
    \node at (\xC + 0.2*\r, \yC-0.2*\r) {$\bm{r}_{c, n}$};
    \node at (\xC - 0.95*\r, \yC+0.5*\r) {$R_n$};

    \fill (\xC, \yC+\R) circle[radius=2pt];
    \node at (\xC, \yC+\R+0.2) {$\bm{r}_{\text{int},n}$};
    
    \draw (\xC, \yC) -- ({\xL + \r*cos(\omegaRad r)}, {\yL + \r*sin(\omegaRad r)});
    \draw[dashed, thin] ({\xL + \r*cos(\omegaRad r)}, {\yL + \r*sin(\omegaRad r)}) arc (145:-190:\R);
    \draw ({\xL + \r*cos(\omegaRad r)}, {\yL + \r*sin(\omegaRad r)}) arc (145:35:\R);

    \draw[thick, ->] ({\xL + \r*cos(\omegaRad r)}, {\yL + \r*sin(\omegaRad r)}) -- ({\xL + \r*cos(\omegaRad r) - 1.5*sin(\omegaRad r)}, {\yL + \r*sin(\omegaRad r) + 1.5*cos(\omegaRad r)});
    \node at (0.7*\r, 1.22*\r) {$\bm{\hat{t}}_{\text{obs},n}$};

    \draw[thick, ->] ({\xL + \r*cos(\omegaRad r)}, {\yL + \r*sin(\omegaRad r)}) -- ({\xL + \r*cos(\omegaRad r) + 1.5*sin(7*pi/36 r)}, {\yL + \r*sin(\omegaRad r) + 1.5*cos(7*pi/36 r)});
    \node at (1.45*\r, 1.15*\r) {$\bm{\hat{t}}_{\text{int},n}$};

    \draw ({\xL + \r*cos(\omegaRad r) - 0.6*sin(\omegaRad r)}, {\yL + \r*sin(\omegaRad r) + 0.6*cos(\omegaRad r)}) arc (115:55:0.6);
    \node at (0.95*\r, 0.85*\r) {$\theta_0$};

    \draw[<->] (\xL + \r, \yL) -- (\xR - \r, \yL);
    \node at (\xC, \yL+0.15*\r) {$d_n$};    

    \fill[black!80] (\xL, \yL) circle [radius=2pt];
    \node at (\xL, \yL-0.2*\r) {$\bm{r}_{\text{obs},n}$};

    \fill[black!80] (\xR, \yR) circle [radius=2pt];
    \node at (\xR, \yR-0.2*\r) {$\bm{r}_{\text{obs},n+1}$};

\end{tikzpicture}
    \caption{The interface between the two fluids in the $n$th opening of the cluster with obstacle locations $\bm{r}_{\text{obs}, n}$ and $\bm{r}_{\text{obs}, n+1}$, and local coordinate system $\{ \bm{\hat{t}}_n, \bm{\hat{n}}_n \}$. The interface is assumed to be circular with radius $R_n$ and center $\bm{r}_{c, n}$; the interface's midpoint is denoted by $\bm{r}_{\text{int},n}$. At the point of contact between the left obstacle and the interface, their tangent vectors are denoted by $\bm{\hat{t}}_{\text{obs},n}$ and $\bm{\hat{t}}_{\text{int},n}$, respectively, and the angle between these vectors is the contact angle, $\theta_0$. The angle $\omega_n$ is measured positively upwards as the angle between $\bm{\hat{t}}_n$ and the line from the left obstacle's center to its point of contact with the interface; the angle $\Omega_n$ is measured positively upwards as the angle between $-\bm{\hat{t}}_n$ and the line from the interface's center to its point of intersection with the left obstacle. The inter-obstacle distance is denoted by $d_n$.}
    \label{fig:poreOpening2D}
\end{figure}
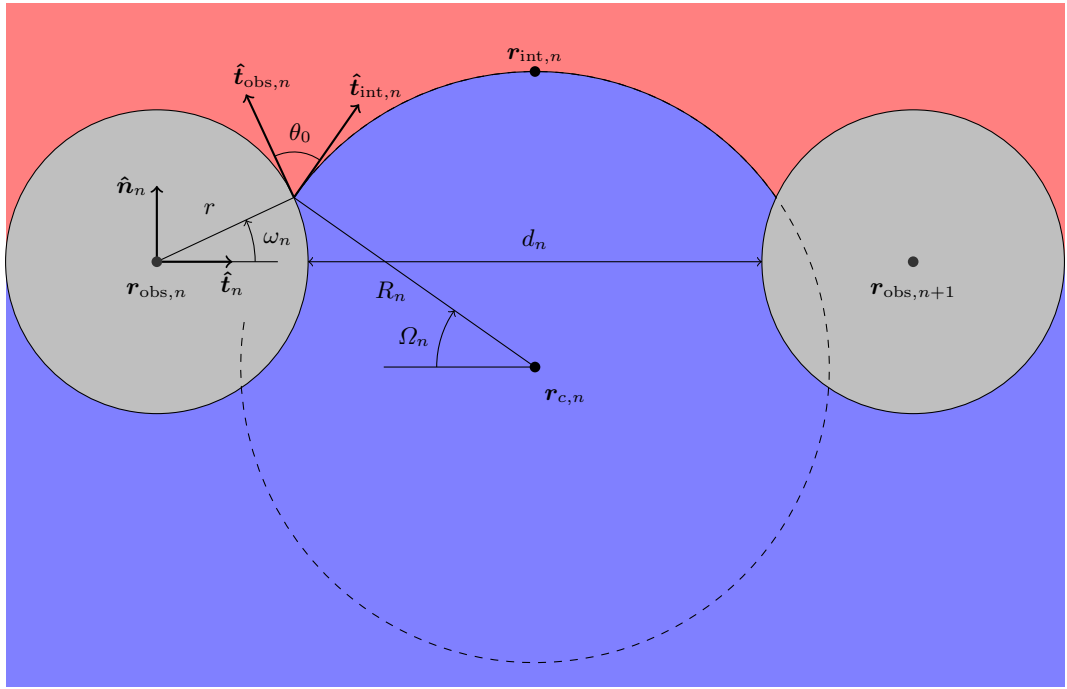

Consider the cluster shown in Figure \ref{fig:clusterParameterization} and its $n$th opening illustrated in Figure \ref{fig:poreOpening2D}. Under the assumptions that the interfaces between the two fluids are circular and infinitely thin, it turns out that the geometry of the cluster is completely determined by the inter-obstacle distances $\bm{d} = [d_1, d_2, ..., d_N]$, the orientation angles $\bm{\xi} = [\xi_1, \xi_2, ..., \xi_N]$ and the opening angles $\bm{\omega} = [\omega_1, \omega_2, ..., \omega_N ]$, where $N$ is the number of cluster openings; figures \ref{fig:clusterParameterization} and \ref{fig:poreOpening2D} define the variables $d_n$, $\xi_n$ and $\omega_n$. Since both $\bm{d}$ and $\bm{\xi}$ are fixed quantities, any change in the cluster's geometry can be described by a change in $\bm{\omega}$ up to the point where an interface collapses and the cluster reconfigures its shape.

The shape of the cluster changes with the Bond number, and in this section we derive a simple, yet accurate model that predicts this change of shape. More specifically, we show that, when all other parameters than the Bond number are held constant, $\bm{\omega}$ satisfies a system of coupled differential equations of the form 
\begin{equation}\label{eq:simplifiedModel}
    \mathsfbi{A} \frac{\partial \bm{\omega}}{\partial (\text{Bo})} = \bm{c},
\end{equation}
where $\mathsfbi{A}$ is a matrix of size $N \times N$ and $\bm{c}$ is a vector of size $N$. Since these quantities depend on $\bm{\omega}$ in complicated, non-linear ways, the system of equations cannot be solved analytically in general, but the time it takes to solve them numerically is substantially smaller than the time it would take to carry out corresponding simulations based on, for example, solving the Navier-Stokes equations for the two phases. In addition, and perhaps more importantly, the derivation of \eqref{eq:simplifiedModel} provides insight into the physical mechanisms that govern the stability of the cluster; in fact, the equations are a consequence of requiring 1) that mass is conserved as $\text{Bo}$ changes, and 2) that the cluster is in mechanical equilibrium up to the point of breaking. We note that 2) corresponds to the assumption that the Bond number is changed over a time scale which is much longer than the time the cluster needs to adjust itself to the new conditions. This is the reason why time does not play any role in \eqref{eq:simplifiedModel} and the reason we call it quasistatic.

In what follows, we first show that the quantities $\bm{d}$, $\bm{\xi}$ and $\bm{\omega}$ indeed determine the geometry of the cluster. Next, we calculate the area of the cluster; as we have assumed the fluids to be incompressible, mass conservation is equivalent to area conservation. We then derive a relation between the curvatures, vertical coordinates and pressures just outside two different cluster openings in mechanical equilibrium, which we use to establish \eqref{eq:simplifiedModel}. We then discuss the stability of a single opening, and this naturally leads to a method for determining a cluster's critical Bond number using this quasistatic model. Finally, we discuss how the change in cluster geometry can be predicted when other parameters than the Bond number are varied.

\subsection{The geometry of a trapped cluster}
When describing a cluster's geometry, it is convenient to label the obstacles in counterclockwise order as shown in Figure \ref{fig:clusterParameterization}, such that the $n$th opening of the cluster lies between the $n$th and the $(n+1)$th obstacle. We equip each opening with a local coordinate system $\{ \bm{\hat{t}}_n, \bm{\hat{n}}_n \}$, where $\bm{\hat{t}}_n$ is the unit vector pointing from the center of the $n$th obstacle, $\bm{r}_{\text{obs}, n}$, towards the center of the $(n+1)$th obstacle, $\bm{r}_{\text{obs},n+1}$, and $\bm{\hat{n}}_n$ is found by rotating $\bm{\hat{t}}_n$ 90 degrees counterclockwise. The orientation of the $n$th local coordinate system relative to the global $\{ \bm{\hat{x}}, \bm{\hat{y}}\}$ coordinate system is encoded in the orientation angle $\xi_n$ between the global $x$-axis and $\bm{\hat{t}}_n$. We emphasize that all these quantities are independent of $\text{Bo}$ and can be calculated once and for all before solving \eqref{eq:simplifiedModel}.

Now, assuming the fluid interfaces to be circular, each of them can be described by their radius, $R_n$, and their center, $\bm{r}_{c,n}$ -- see Figure \ref{fig:poreOpening2D}. One way of calculating these quantities in terms of the opening angle $\omega_n$ is to parameterize the curves along the circumference of the $n$th obstacle and the $n$th interface; the two curves should intersect when the angles of the parameterizations are $\omega_n$ and $\Omega_n$, respectively, and at the intersection point the angle between the two curves should be the contact angle, $\theta_0$. In the local coordinate system, a parameterization of the circumference of the $n$th obstacle is given by
\begin{equation}\label{eq:localObstacleParameterization}
    \big( x(\omega), y(\omega) \big) = \bm{r}_{\text{obs},n} + r \cos(\omega) \bm{\hat{t}}_n + r \sin(\omega) \bm{\hat{n}}_n, 
\end{equation}
while the $n$th fluid interface can be parameterized as
\begin{equation}\label{eq:localInterfaceParameterization}
    \big( x(\Omega), y(\Omega) \big) = \bm{r}_{c,n} - R_n \cos(\Omega) \bm{\hat{t}}_n + R_n \sin(\Omega) \bm{\hat{n}}_n. 
\end{equation}
We note that the obstacle parameterization runs counterclockwise and that the interface parameterization runs clockwise, in accordance with the signs of $\omega_n$ and $\Omega_n$ defined in Figure \ref{fig:poreOpening2D}. The angle $\Omega_n$ is found by requiring that the angle between the two curves' tangent vectors $\bm{\hat{t}}_{\text{obs},n}$ and $\bm{\hat{t}}_{\text{int},n}$ should equal the contact angle, and a straightforward calculation shows that
\begin{equation}\label{eq:OMEGA}
    \Omega_n = \theta_0 - \omega_n.
\end{equation}
In order to calculate the radius of the fluid interface, we note that symmetry implies that its center must lie halfway between the obstacles, and hence that $(\bm{r}_{c,n} - \bm{r}_{\text{obs},n})\cdot \bm{\hat{t}}_n = r + d_n/2$. Equating \eqref{eq:localObstacleParameterization} and \eqref{eq:localInterfaceParameterization} at the point where $\omega = \omega_n$ and $\Omega = \Omega_n$ and taking the dot product with $\bm{\hat{t}}_n$ therefore implies that the normalized interface radius is given by the expression  
\begin{equation}\label{eq:interfaceRadius}
    \frac{R_n}{r} = \frac{d_n/r + 2\big( 1 - \cos(\omega_n) \big) }{2 \cos(\Omega_n)}.
\end{equation}
The geometric description of the interface is completed by equating \eqref{eq:localObstacleParameterization} and \eqref{eq:localInterfaceParameterization} at the intersection point and taking the dot product with $\bm{\hat{n}}_n$; when combined with the symmetry argument above this operation yields
\begin{equation}\label{eq:interfaceCenter}
    \frac{\bm{r}_{c,n}}{r} = \frac{\bm{r}_{\text{obs},n}}{r} + \bigg(1 + \frac{d_n}{2 r} \bigg) \bm{\hat{t}}_n - \bigg( \frac{R_n}{r}\sin(\Omega_n) - \sin(\omega_n) \bigg) \bm{\hat{n}}_n.
\end{equation}

\subsection{The area of a trapped cluster}
In order to derive the area of a trapped cluster, we again consider the cluster depicted in Figure 
\ref{fig:clusterParameterization}. Using Green's theorem, its area, $A_{\text{cluster}}$, can be calculated as the line integral 
\begin{equation}\label{eq:greensTheorem}
    A_{\text{cluster}} = \frac{1}{2} \oint_{\mathcal{C}} \big( -y \, \diff x + x \, \diff y\big),
\end{equation}
where $\mathcal{C}$ is the closed curve along the cluster's circumference oriented counterclockwise. We note that the integral only equals the cluster area in the case where the cluster contains no embedded obstacles, that is, obstacles that are completely surrounded by wetting fluid, which may be present in the case of large clusters. However, since our only application of the cluster area is to make sure that it remains constant when the Bond number changes, the constant contributions to the area from embedded obstacles can be neglected for the present purpose. To make progress from here, we decompose $\mathcal{C}$ into contributions from the fluid interfaces and the obstacles as shown in Figure \ref{fig:clusterParameterization}, which gives that
\begin{equation}\label{eq:contourDecomposition}
    A_{\text{cluster}} = 
    \frac{1}{2} \sum_{n = 1}^{N}
    \bigg( \oint_{\mathcal{C}_{\text{int},n}} \big( -y \, \diff x + x \, \diff y\big) 
    + 
    \oint_{\mathcal{C}_{\text{obs},n}} \big( -y \, \diff x + x \, \diff y\big)
    \bigg). 
\end{equation}
To calculate the contribution from the $n$th fluid interface, we employ the parameterization \eqref{eq:localInterfaceParameterization} with $\Omega_n \le s \le \pi - \Omega_n$. Carrying out the integral yields the expression
\begin{equation}\label{eq:areaInterfaceContribution}
    \int_{\mathcal{C}_{\text{int},n}} \big( -y \, \diff x + x \, \diff y \big)
        = - 2 R_n \cos \big(\Omega_n \big) \big( \bm{r}_{\text{int},n} \cdot \bm{\hat{n}}_n \big)
        -  R_n^2 \big( \pi - 2\Omega_n \big),
\end{equation}
in which $\bm{r}_{\text{int},n} \equiv \bm{r}_{c,n} + R_n \bm{\hat{n}}_n$ is the midpoint of the $n$th interface as illustrated in Figure \ref{fig:poreOpening2D}. For the integral along the obstacle circumference, instead of using the local coordinate system, it turns out to be beneficial to use the global $\{ \bm{\hat{x}}, \bm{\hat{y}}\}$-coordinate system to parameterize the curve segment. One possible choice is
\begin{equation}\label{eq:obstacleParameterization}
    \big( x(s), y(s) \big)
    = 
    \bm{r}_{\text{obs},n} - r \cos(s) \bm{\hat{x}} + r \sin(s) \bm{\hat{y}}, 
    \qquad
    \text{where}
    \qquad
    s_{1, n} \le s \le s_{2, n},
\end{equation}
in which the upper limit of the parameterization is $s_{2,n} = \pi - \xi_n - \omega_n$ while the lower limit is $s_{1,n} = 2\pi - \xi_{n-1} + \omega_{n-1}$ provided that this value is smaller than $s_{2,n}$; if it is larger than $s_{2,n}$, $2\pi$ should be subtracted from $s_{1,n}$. This technicality only matters when evaluating the term $r^2 (s_{2,n} - s_{1,n})$ in \eqref{eq:areaObstacleContribution} below, and we note that the subtraction of $2\pi$ is in fact irrelevant when establishing \eqref{eq:simplifiedModel}, since it disappears when differentiating with respect to $\text{Bo}$. Inserting the parameterization \eqref{eq:obstacleParameterization} into the line integral, the contribution from the $n$th obstacle segment turns out to be
\begin{equation}\label{eq:areaObstacleContribution}
    \begin{split}
        \int_{\mathcal{C}_{\text{obs}},n} \big( -y \, \diff x + x \, \diff y \big)
        & = r y_{\text{obs},n} \big( \cos(s_{2,n}) - \cos(s_{1,n}) \big) \\
        & + r x_{\text{obs},n} \big( \sin(s_{2,n}) - \sin(s_{1,n}) \big) \\
        & - r^2 (s_{2,n} - s_{1,n}),
    \end{split}
\end{equation}
where $x_{\text{obs},n}$ and $y_{\text{obs},n}$ are the $x$- and $y$-coordinates of $\bm{r}_{\text{obs},n}$, respectively. In combination the results \eqref{eq:contourDecomposition}, \eqref{eq:areaInterfaceContribution} and \eqref{eq:areaObstacleContribution} imply that the cluster area must satisfy the relation
\begin{equation}\label{eq:clusterArea2}
    \begin{split}
    2A_{\text{cluster}}/r^2 =  \sum_{n = 1}^{N} f_n (\omega_n, \omega_{n-1}),
    \end{split}
\end{equation}
in which the function $f_n(\omega_n, \omega_{n-1})$ is defined as
\begin{equation}\label{eq:nthClusterArea}
    \begin{split}
        f_n(\omega_n, \omega_{n-1})
        \equiv
        & - \frac{2 R_n}{r} \cos\big(\Omega_n \big) \big( (\bm{r}_{c, n} / r) \cdot \bm{n}_n \big) 
        - \bigg(\frac{R_n}{r} \bigg)^2 \big( \pi - 2\Omega_n \big) \\
        & +  
        \frac{y_{\text{obs},n}}{r} \big( \cos(s_{2,n}) - \cos(s_{1,n}) \big) \\
        & + \frac{x_{\text{obs},n}}{r} \big( \sin(s_{2,n}) - \sin(s_{1,n}) \big) \\
        & - (s_{2,n} - s_{1,n}).
    \end{split}
\end{equation}
This completes the calculation of the cluster area.

\subsection{The pressure difference across a trapped cluster}\label{subsec:pressureCurvatureRelation}
As long as a cluster is trapped, the pressure just outside its $n$th opening is related to the pressure just outside its $i$th opening. To show this, we first define the outer and inner points of the $n$th fluid interface as $\bm{r}_{\text{out},n} \equiv \bm{r}_{\text{int},n} - \delta \bm{\hat{n}}_n$ and $\bm{r}_{\text{in},n} \equiv \bm{r}_{\text{int},n} + \delta \bm{\hat{n}}_n$, and equivalently for the $i$th interface. Here, $\delta$ is a length scale which is very large relative to the width of the interface and very small relative to the obstacle radius. Next, we consider the momentum balance across the $n$th interface, which requires that the equation
\begin{equation}\label{eq:interfaceMomentumBalance}
\begin{split}
    \big( p_{\text{out},n} - p_{\text{in}, n} \big) \bm{\hat{n}}_n
    = &
    \gamma \kappa_n \bm{\hat{n}}_n
    + \mu_\text{nw} \Big(\nabla \bm{u}_{\text{out},n} + (\nabla \bm{u}_{\text{out},n})^\top \Big) \bm{\hat{n}}_n \\
    & - \mu_\text{w} \Big(\nabla \bm{u}_{\text{in},n} + (\nabla \bm{u}_{\text{in},n})^\top \Big) \bm{\hat{n}}_n
\end{split}
\end{equation}
is satisfied (see e.g. the derivation in section 1.1.2 in the book by \cite{GrossReusken2011}). In this equation, $\kappa_n \equiv 1/R_n$ is the curvature of the $n$th interface, $\bm{u}$ is the velocity field, and the subscripts ``out,$n$'' and ``in,$n$'' signify that the sub-scripted quantity is evaluated at $\bm{r}_{\text{out}, n}$ and $\bm{r}_{\text{in}, n}$, respectively. Since $\bm{u}$ is continuous across the interface, we note that the fluid inside the trapped cluster may in fact circulate. From the simulations presented in Sections \ref{sec:validation} and \ref{sec:stabilityProb} we have, however, found that the fluid velocity and its gradient inside the trapped cluster are always very small relative to the corresponding quantities of the outside flow, and for that reason we neglect the last term on the right-hand side of \eqref{eq:interfaceMomentumBalance}. Interestingly, this implies that the viscosity of the wetting fluid does not impact the stability of the cluster, and, as a consequence, that the ratio of the two fluids' viscosities cannot be a relevant parameter for cluster stability (provided, as in this case, that substantial circulation inside the cluster can be ruled out). In addition, a crude order-of-magnitude estimate shows that $\mu_\text{nw} |\nabla \bm{u}_{\text{out},n} | /(\gamma \kappa_n) = O(\text{Ca})$, where $\text{Ca} \equiv \mu_\text{nw} v_0 / \gamma$ is the capillary number of the flow in the non-wetting fluid and $v_0$ is a typical velocity of the flow. Hence, as long as the capillary number is small, the second term on the right-hand side of \eqref{eq:interfaceMomentumBalance} can be neglected and the pressure differences across the $n$th and $i$th interfaces are given by the well-known Young-Laplace equation, i.e.
\begin{equation}\label{eq:interfacePressureDiff}
\left\{
\begin{aligned}
p_{\text{out},n} - p_{\text{in},n} & = \kappa_n \gamma, \\
p_{\text{out},i} - p_{\text{in},i} & = \kappa_i \gamma.
\end{aligned}
\right.
\end{equation}
As is evident from these equations, we choose the curvature to be positive when the pressure on the outside is larger than on the inside, that is, $\kappa_n$ is positive whenever the non-wetting side of the interface is convex. The final step in relating $p_{\text{out},n}$ to $p_{\text{out},i}$ consists in noting that the pressure inside the trapped cluster must be hydrostatic, because the fluid--to a very good approximation--lies still there. By this we mean that we must have
\begin{equation}\label{eq:hydrostaticPressure}
    p_{\text{in},n} - p_{\text{in},i} 
    =
    \rho_{\text{w}} g (y_{\text{int}, n} -  y_{\text{int,i}}),
\end{equation}
and if we subtract the two equations in \eqref{eq:interfacePressureDiff}, use \eqref{eq:hydrostaticPressure} to eliminate the pressure difference inside the cluster and normalize by $\rho_{\text{w}} g r$, we find that
\begin{equation}\label{eq:pressureCurvatureRelation}
    \frac{p_{\text{out},n} - p_{\text{out},i}}{\rho_{\text{w}} g r}
    = 
    \frac{\kappa_n r - \kappa_i r}{\text{Bo}}
    + 
    \frac{y_{\text{int},n} - y_{\text{int},i}}{r}.
\end{equation}
At a first glance, one would think that this relation imposes $N (N - 1)/2$ constraints on the geometry of the cluster, because it should hold for all possible combinations of $n$ and $i$. However, if it holds for $i = 1$ and all $n \ge 2$, then the relation is automatically satisfied for all possible choices of $n$ and $i$. This is implied by the linearity of the relation. As such, \eqref{eq:pressureCurvatureRelation} only imposes $N-1$ constraints on the cluster geometry.

\begin{figure}
    \centering
    \includegraphics[width=1.0\textwidth]{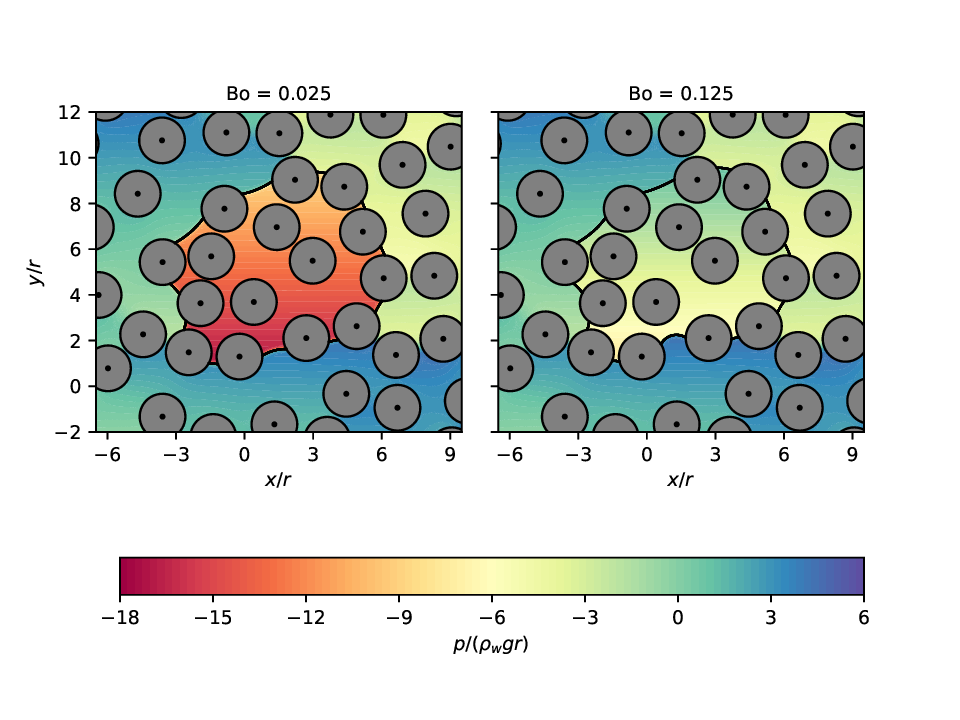}
    \caption{The normalized pressure field around a trapped cluster for $\text{Bo} = 0.025$ and $\text{Bo} = 0.125$. In the two cases, the cluster is either very far (left figure) of very close (right figure) to breaking, respectively, yet the pressure fields around the cluster are essentially the same. The pressure inside the cluster is larger in the small-$\text{Bo}$ case than in the large-$\text{Bo}$ case, because the Bond number is varied by varying the surface tension while keeping the gravitational acceleration constant. The pressure field is calculated using the DNS model described in section \ref{subsec:fullModel} using a circular cluster with radius $r_\text{circle}/r = 4$ as the initial condition.}
    \label{fig:constantPressureBeforeBreaking}
\end{figure}

In what follows, we will assume that the left hand side of \eqref{eq:pressureCurvatureRelation} is independent of the Bond number. As we envision changing the Bond number by varying either $g$ or $\gamma$, this is reasonable when recalling that a creeping flow is accurately described by the Stokes equation and therefore linear in the driving force and otherwise determined by the geometry. Since the pressure difference is normalized by $g$, changing the value of this parameter does not have a direct effect on the left hand side. Moreover, as we only consider the range $\text{Bo} \le \text{Bo}_\text{crit}$, the changes in the cluster geometry are necessarily small. In particular, no pathways for the non-wetting fluid are created nor destroyed in this range. As such, the geometrical contribution to the change in the normalized pressure difference is expected to be small, which it indeed is. Figure \ref{fig:constantPressureBeforeBreaking} shows an example of this by comparing a trapped clusters in two different states in which it is either very far or very close to breaking. The validation of the quasistatic model against numerical simulations in Section \ref{sec:validation} provides further evidence that the assumption is indeed satisfied to a good approximation.

\subsection{The system of differential equations governing the cluster geometry}
Equations \eqref{eq:clusterArea2} and \eqref{eq:pressureCurvatureRelation} govern the geometry of the cluster as the Bond number changes. As we have shown, all geometric changes are ultimately determined by changes in the opening angles, $\bm{\omega} = [\omega_1, \omega_2, ..., \omega_N]$, which in turn depend on a long list of parameters. However, as we are presently only concerned with changes in the opening angles induced by changes in $\text{Bo}$, we consider $\omega_n$ a function of this variable only. Hence, if we require the mass of the cluster to be independent of the Bond number, taking the derivative of \eqref{eq:clusterArea2} with respect to $\text{Bo}$ yields
\begin{equation}\label{eq:clusterArea3}
    0 = \sum_{n = 1}^{N} \bigg( \frac{\partial f_n}{\partial \omega_n} + \frac{\partial f_{n+1}}{\partial \omega_n} \bigg) \frac{\partial \omega_n}{\partial (\text{Bo})}.
\end{equation}
Here periodic indexing is employed such that $f_{N+1}(\omega_{N+1}, \omega_{N}) = f_1(\omega_1, \omega_{N})$. This is the first of the $N$ equations that constitute \eqref{eq:simplifiedModel}; the remaining $N-1$ equations are derived from the relation \eqref{eq:pressureCurvatureRelation}. Setting $i = 1$ and making the assumption that the left hand side is independent of the Bond number, taking the derivative with respect to $\text{Bo}$ yields
\begin{equation}\label{eq:pressureConservation}
    \bigg( \frac{\diff (y_{\text{int},1}/r)}{\diff \omega_1}  + \frac{1}{\text{Bo}} \frac{\diff (\kappa_1 r)}{\diff \omega_1} \bigg) \frac{\partial \omega_1}{\partial (\text{Bo})}
    - 
    \bigg( \frac{\diff (y_{\text{int},n}/r)}{\diff \omega_n} + \frac{1}{\text{Bo}} \frac{\diff (\kappa_n r)}{\diff \omega_n} \bigg) \frac{\partial \omega_n}{\partial (\text{Bo})}
    = 
    \frac{\kappa_1 r - \kappa_n r}{(\text{Bo})^2},
\end{equation}
for $n = 2, 3, ..., N$. In combination, \eqref{eq:clusterArea3} and \eqref{eq:pressureConservation} constitute the system of $N$ coupled equations \eqref{eq:simplifiedModel} governing the evolution of the opening angles as a function of the Bond number. To arrive at the matrix formulation \eqref{eq:simplifiedModel}, we simply define
\begin{subequations}\label{eq:matrixElements}
\begin{align}
   a_n & \equiv \frac{\partial f_n}{\partial \omega_n}  + \frac{\partial f_{n+1}}{\partial \omega_n}, \\
   b_n & \equiv \frac{\diff (y_{\text{int},n}/r)}{\diff \omega_n} + \frac{1}{\text{Bo}}\frac{\diff (\kappa_n r)}{\diff \omega_n}, \\
   c_n & \equiv \frac{\kappa_1 r - \kappa_n r }{(\text{Bo})^2},
\end{align}
\end{subequations}
and set the matrix $\mathsfbi{A}$ and vector $\bm{c}$ to
\begin{equation}\label{eq:conservationMatrix}
    \mathsfbi{A} = 
    \begin{bmatrix}
        a_1    & a_2    & a_3    & \cdots & a_{N-1}  & a_{N} \\
        b_1    & -b_2   & 0      & \cdots & 0                    & 0 \\
        b_1    & 0      & -b_3   & \cdots & 0                    & 0 \\
        \vdots & \vdots & \vdots & \ddots & \vdots               & \vdots \\
        b_1    & 0      & 0      & \cdots & -b_{N-1} & 0 \\
        b_1    & 0      & 0      & \cdots & 0                    & -b_{N} \\
    \end{bmatrix}
    , \qquad \qquad \qquad
    \bm{c} = 
    \begin{bmatrix}
        0   \\
        c_2 \\
        c_3 \\
        \vdots \\
        c_{N-1} \\
        c_{N}
    \end{bmatrix}.
\end{equation}
Given an initial configuration, $(\bm{\omega}_0, \text{Bo}_0)$, the set of equations \eqref{eq:simplifiedModel} is easily integrated numerically with respect to the Bond number by any standard ODE solver. As such, it only remains for us to clarify, up to which Bond number one should integrate in order to predict the point of breaking. This is done in the following.

\subsection{When does a cluster become unstable?}
\begin{figure}
    \centering
    \includegraphics[width=0.7\textwidth]{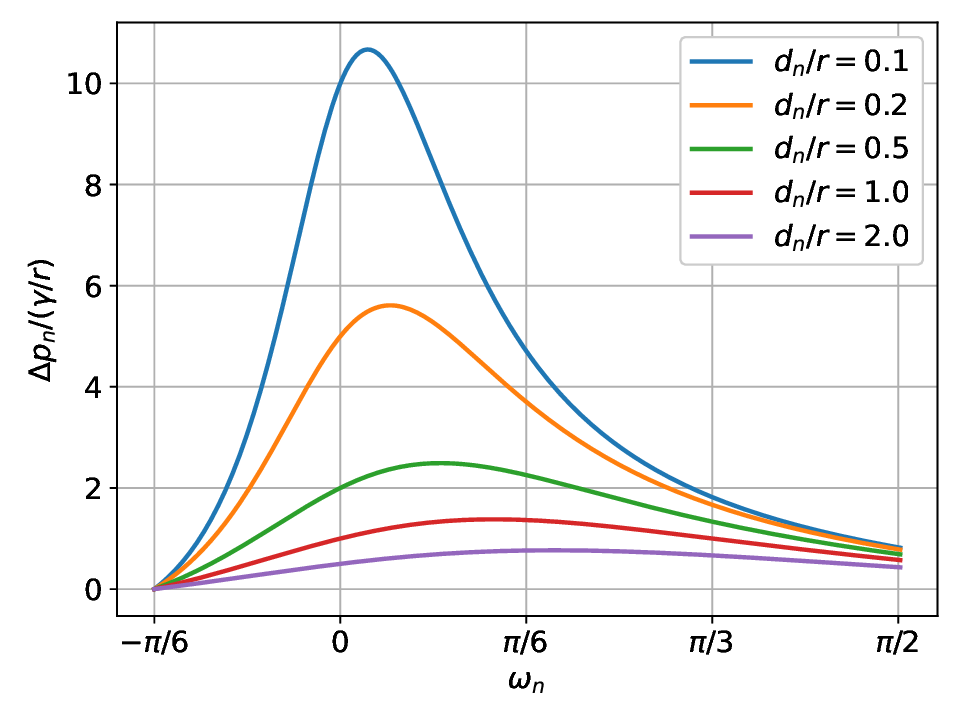}
    \caption{The pressure difference across the $n$th cluster opening given by \eqref{eq:youngLaplace2} as a function of the angle $\omega_n$ for $\theta = \pi/3$ and different values of the non-dimensional inter-obstacle distance $d_n/r$.}
    \label{fig:pressureDifferenceVsAngle}
\end{figure}

As long as the Bond number is smaller than its critical value, the interfaces of the cluster can balance out the pressure difference across them by adjusting their shapes. More precisely, if the pressure difference across the $n$th interface is $\Delta p_n$, then the interface will adjust its curvature such that the Young-Laplace equation is satisfied. Using the expression \eqref{eq:interfaceRadius} for the radius of the interface, the expression \eqref{eq:OMEGA} for the angle $\Omega_n$ and normalizing by $\rho_\text{w} g r$, the Young-Laplace equation can be written as
\begin{equation}\label{eq:youngLaplace2}
    \frac{\Delta p_n}{\rho_\text{w} g r} = 
    \frac{1}{\text{Bo}} \, 
    \frac{2 \cos(\theta_0 - \omega_n)}{d_n/r + 2\big( 1 - \cos(\omega_n) \big)}.
\end{equation}
This is the (normalized) pressure difference that the $n$th interface can sustain by opening up to the angle $\omega_n$. Plots of this pressure difference as a function of $\omega_n$ are shown in Figure \ref{fig:pressureDifferenceVsAngle} for $\theta_0 = \pi/3$ for different values of $d_n/r$, and from this it is clear that the pressure difference has a global maximum. One may interpret this result in the following way: if the cluster is initially stable, it can respond to an increasing pressure on its outer side, a decreasing pressure on its inner side or a weakening of the surface tension by moving inwards. Since the contact angle is constant, the new position further inwards will ensure that the curvature of the interface is larger and, therefore, that the interface becomes stronger. This, however, only works up to a certain, critical angle which we denote by $\omega_{\text{crit},n}$. Beyond this angle, further inwards movement will result in a smaller curvature and hence a weakening of the interface. Differentiating \eqref{eq:youngLaplace2} with respect to $\omega_n$, it is straightforward to show that the critical angle is given by the expression
\begin{equation}\label{eq:omegaCrit}
    \omega_{\text{crit},n} = \theta_0 - \sin^{-1} \bigg( \frac{2 \sin(\theta_0)}{d_n/r + 2} \bigg).
\end{equation}
When the flow conditions are such that one of the cluster's openings is pushed to its critical value, the cluster becomes unstable. For if some small perturbation pushes this interface just a tiny bit further in, it will become weaker and therefore keep yielding to the outside pressure until it finally collapses. The critical value of the Bond number for a given cluster can thus be computed by solving \eqref{eq:simplifiedModel} up to the point where $\omega_n = \omega_{\text{crit},n}$ for some $n$.

We note that this instability is the well-known Haines jump, discovered by \cite{Haines1930} now nearly a century ago. \cite{CieplakRobbins1988} refer to this event as a ``burst'', and as they have shown, there are other ways in which an interface can be unstable. For example, an interface may split into two new interfaces if it touches another obstacle before reaching its critical angle. The detection of such an event could in principle be built into the numerical solution of the model proposed here. We have, however, not done so because the continuation after the touch would most likely require a full (highly expensive) recomputation of the outside pressure field. While the small movements of an interface do not lead to large changes in the surrounding pressure, a reconfiguration of the interfaces oftentimes does. For small clusters we note that this simplification does not affect the computation of $\text{Bo}_\text{crit}$, because these do not have any embedded obstacles such that a touching event could take place. For large clusters, however, we acknowledge that the simplification may add to the uncertainty of the computed values of $\text{Bo}_\text{crit}$.

\subsection{Varying other parameters than the Bond number}
The Bond number is not the only parameter on which the trapped cluster's geometry depends; for example, it also depends on the ratio of fluid densities, $\rho_\text{w}/\rho_\text{nw}$, and the contact angle, $\theta_0$, and using an approach similar to the one above, it is possible to derive equations corresponding to \eqref{eq:simplifiedModel} for predicting the change in $\bm{\omega}$ when these parameters are varied. In Appendix \ref{app:simplifiedModelRho} we show and validate that when the ratio $\rho_\text{w}/\rho_\text{nw}$ is varied while keeping all other variables fixed, the opening angles satisfy the system of equations
\begin{equation}\label{eq:simplifiedModelRho}
    \mathsfbi{A} \frac{\partial \bm{\omega}}{\partial (\frac{\rho_\text{w}}{\rho_\text{nw}})} = \bm{c}_{\rho_\text{w}/\rho_\text{nw}},
\end{equation}
in which $\mathsfbi{A}$ is the same as in \eqref{eq:simplifiedModel} and $\bm{c}_{\rho_\text{w}/\rho_\text{nw}}$ is given by \eqref{eq:cRho}. We note that this set of equations allows for cost-efficient computation of the critical Bond number as a function of the density ratio. Given a cluster geometry for particular values of $\text{Bo}$ and $\rho_\text{w}/\rho_\text{nw}$, one can first integrate $\bm{\omega}$ with respect to the density ratio up to the desired value using \eqref{eq:simplifiedModelRho} before computing $\text{Bo}_\text{crit}$ using \eqref{eq:simplifiedModel}. A set of equations that allows for changes in $\theta_0$ can likewise be derived, but we have not done so in the present work. A quasistatic model for the change in cluster geometry with respect to the parameters of the porous geometry can, however, not be derived. Studying how $\text{Bo}_\text{crit}$ depends on, say, the minimum required distance between two obstacles, would require completely new initial conditions for the trapped cluster geometries.

At this point, a comment on computational efficiency is appropriate, as we have repeatedly asserted that the quasistatic model is significantly faster than direct numerical simulations for predicting the critical Bond number. This claim may appear counterintuitive, given that the quasistatic model currently requires an initial condition that can only be obtained through a direct simulation. The key advantage, however, lies in the fact that only a single expensive simulation per cluster is needed to compute the critical Bond number as a function of the density ratio and, potentially, the contact angle. In contrast, without the quasistatic model, 5–10 direct simulations would be required for each combination of density ratio and contact angle. For a single pair of the density ratio and the contact angle, the speed-up is therefore roughly 5 to 10 times, while, if a detailed study across $(\rho_\text{w}/\rho_{nw}, \theta_0)$-space was to be carried out, it is fair to say that employing the quasistatic model would lead to computational savings of several orders of magnitude.

In this connection we note that the full numerical simulations based on phase-field modeling presented in Sections \ref{sec:validation} and \ref{sec:stabilityProb} are most likely not the most efficient way of initializing the quasistatic model. In the limit of small Bond number (that is, when surface tension is much more decisive for the shape of the cluster than the pressure gradient of the surrounding flow), the equilibrium configuration of the cluster surfaces could possibly be obtained directly, for example, by the Surface Evolver software of \cite{Brakke1992}. The surrounding pressure field could subsequently be computed by solving the Stokes equation once for the entire domain employing the cluster boundaries found by the surface evolver. In the present work we have, however, not tested this strategy, and more research is needed to clarify how the initialization of the quasistatic model can be accelerated.

\section{Validation of the quasistatic model}\label{sec:validation}
In order to validate the quasistatic model \eqref{eq:simplifiedModel}, we compare its predictions to direct numerical simulations (DNS) based on the phase-field model of \cite{DingEtAl2007}. In what follows, we shall refer to this model as ``the DNS model'', and we note that it resolves both dynamic interface motion and the full dynamics of the fluids in the bulk phases. The fundamental quantity of the DNS model is the phase field, $\phi$, which takes the values $1$ and $-1$ in the parts of the domain occupied by the wetting fluid and non-wetting fluid, respectively; for $\phi$-values in between these two extremes, the fluid is a mixture of the two phases. The transition between the two phases occurs over the length scale $\epsilon$ in the direction parallel to $\nabla \phi$. We note that \cite{AbelsEtAl2012} have shown that the model converges to the sharp-interface description in the limit of small $\epsilon$. However, since for any finite value of $\epsilon$ the phase-field model treats the two fluids as one by interpolating their properties based on $\phi$, a numerical implementation of the model does not need to track the interface between the fluids. This is a clear computational advantage compared to the sharp-interface description. Below, we provide a more detailed description of the DNS model, before turning to the validation of the quasistatic model.

\subsection{The DNS model}\label{subsec:fullModel}
The phase-field model for two-phase flow, as formulated by \cite{DingEtAl2007}, consists of a Navier-Stokes-Cahn-Hilliard system of differential equations supplemented with appropriate boundary conditions at solid interfaces. Letting $\mathbf{u}$ be the velocity of the fluids and $p$ be the pressure, the governing equations read
\begin{subequations}\label{eq:fullModel}
    \begin{align}
    \rho(\phi) \left[ \frac{\partial \bm{u}}{\partial t} 
    + \big( \bm{u} \cdot \nabla \big) \bm{u} \right]
    - \nabla \cdot \big( 2 \mu(\phi) \bm{D u} \big)
    + \nabla p
    & = g_\phi \nabla \phi + \rho(\phi) \bm{g}, \label{eq:fullModel_momentum}\\
    \nabla \cdot \bm{u} 
    & = 
    0, \label{eq:fullModel_mass}\\
    \frac{\partial \phi}{\partial t} 
    + \bm{u} \cdot \nabla \phi   
    & =
    \epsilon M_0 \nabla^2 g_\phi, \label{eq:fullModel_phi}\\
    g_\phi 
    & = \frac{3 \gamma}{2 \sqrt{2}} \Big( \epsilon^{-1} W'(\phi) - \epsilon \nabla^2 \phi \Big).
\end{align}
\end{subequations}
Here, $\rho(\phi)$ and $\mu(\phi)$ are linear interpolations of the mass densities and dynamic viscosities of the two fluids, respectively. For example,
\begin{equation}\label{eq:densityInterpolation}
    \rho(\phi) \equiv \frac{1 + \phi}{2} \rho_\mathrm{w} +  \frac{1-\phi}{2}\rho_\mathrm{nw},
\end{equation}
with $\mu(\phi)$ defined completely analogously. Moreover, $\bm{Du} \equiv (\nabla \bm{u} + (\nabla \bm{u})^\top )/2$ is the rate-of-strain tensor, $M_0$ is the phase field mobility and $W(\phi) \equiv ( 1 - \phi^2)^2 /4$ is the Ginzburg-Landau double well potential. It should be noted that \eqref{eq:fullModel} is not strictly thermodynamically consistent; in contrast, the model by \cite{AbelsEtAl2012} obtains thermodynamic consistency by including a small term in \eqref{eq:fullModel_momentum} that makes the model respect a version of the 2nd law of thermodynamics for finite $\epsilon > 0$. However, for small $\epsilon$ we expect little difference between the two formulations as shown numerically by \cite{AlandVoigt2012}.

At solid boundaries, the following conditions apply (see e.g.\ \cite{CarlsonEtAl2012}): 
\begin{subequations}
\begin{align}
    \bm{u} &= \bm{0}, \\
    \hat{\bm{n}} \cdot \nabla g_\phi &= {0}, \\
    \hat{\bm{n}} \cdot \nabla \phi &= \frac{2\sqrt{2}}{3 \epsilon } \cos (\theta_0) f'_w (\phi) \label{eq:fullModel_bc_phi},
\end{align}    
\end{subequations}
where $f_w(\phi) \equiv (2 + 3\phi - \phi^3)/4$ smoothly interpolates between 0 and 1 to enforce the difference in the solid-liquid interface energy between the two liquid phases, and $\hat{\bm n}$ is the normal vector of the solid boundary pointing into the solid. Periodic boundary conditions are applied to all fields on the horizontal and vertical sides of the domain.

We employ the open source finite element code Twoasis (see \cite{Linga2025}) to solve the governing equations of the DNS model numerically. This code is based on the Oasis solver \citep{MortensenValenSendstad2015} built on the open source framework Fenics (see \cite{LoggEtAl2012}). As such, it is capable of discretizing the spatial dimensions of the governing equations using, e.g., Lagrange elements of any desired order. However, since Twoasis carries out the time-integration using a semi-implicit splitting scheme which is first-order accurate in the time step, we use first-order spatial elements in order to treat space and time equally. We generate doubly periodic triangular meshes for the simulations using the built-in meshing tool in the Bernaise solver \citep{LingaEtAl2019}. This tool uses the Python library MeshPy, which is based on the mesh generator Triangle written by \cite{Shewchuk1996}. For further details about the code and its implementation, we refer the reader to the source code, which is freely available from Github at the URLs given in the reference list. The parameter values used in the DNS computations are specified in the next section.

\subsection{Simulation protocol and computational parameters}
We solve the governing equations \eqref{eq:fullModel} in a square domain of size $L \times L$ and start all our simulations with both the velocity field, $\bm{u}$, the pressure, $p$, and the phase field chemical potential, $g_\phi$, equal to zero everywhere in the domain. The values of $\phi$ are chosen such that the domain is initially filled with non-wetting fluid except for a circle of radius $r_\text{circ}$ which contains wetting fluid. Naturally, then, $r_\text{circ}$ becomes the length scale for the size of the trapped cluster. The location of the circle's center is chosen randomly from a uniform distribution over the domain, and at later stages we obtain statistical properties of the cluster stability for a given value of $r_\text{circle}/r$ by carrying out the simulation for an ensemble of random locations of the circle center drawn from a uniform distribution over the domain. In order to treat clusters of different sizes similarly, we choose $L = 8 r_\text{circ}$; in this way, the fraction of the domain which is blocked by the trapped cluster becomes independent of the value of the circle radius.

\begin{figure}
    \centering
    \includegraphics[width=0.95\textwidth]{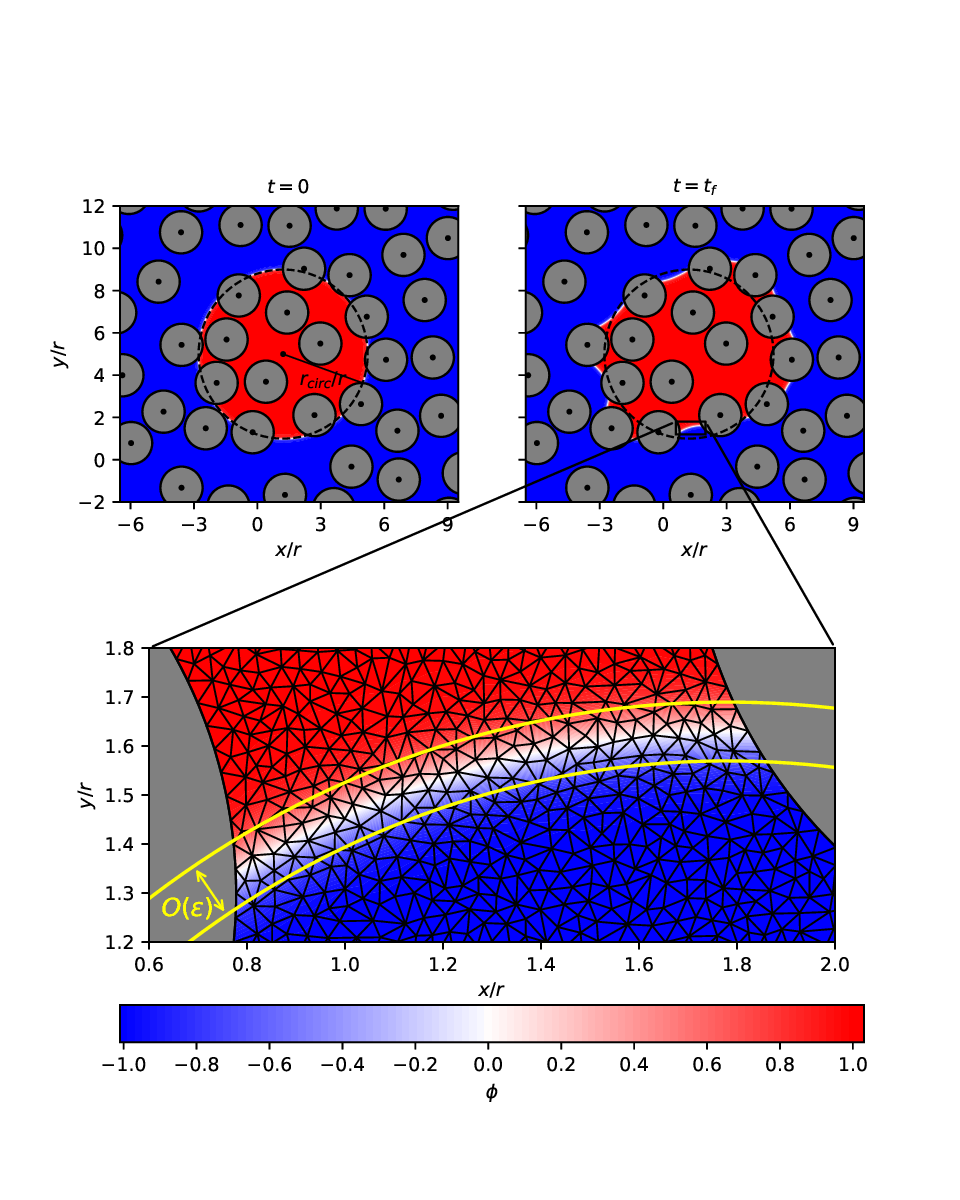}
    \caption{Top row: the initial and final states of a cluster for $r_\text{circ}/r = 4$ and $\text{Bo = 0.025}$. Over time, the wetting fluid is redistributed from the circular configuration to one where the contact angle is everywhere equal to $\theta_0$. Bottom row: a close-up on the diffuse interface, whose width is proportional to $\epsilon$, in one of the cluster openings in the final state together with the employed triangular mesh. As can be seen, the interface is resolved by 3-4 elements.}
    \label{fig:phaseFieldStartAndEnd}
\end{figure}

To enable comparisons, we use the same fluid properties as those used by \cite{MathiesenEtAl2023} in our simulations, that is, we let $\rho_\text{w}/\rho_\text{nw} = 1.25$, $\mu_\text{w}/\mu_\text{nw} = 0.7125$ and $\theta_0 = \pi/3$. These values correspond to a system filled with water and silicone oil. Moreover, we fix the absolute values of $r$, $\rho_\text{w}$, $\mu_\text{w}$ and $g$, and let the width of the phase transition be $\epsilon/r = 0.05$, the mesh size be $h/r = 0.05$ and the phase field mobility be $\epsilon M_0/(\rho_\text{w} (r/g)^{3/2}) = 10^{-11/2}$. As is illustrated in the bottom row of Figure \ref{fig:phaseFieldStartAndEnd}, these choices ensure that the interface is resolved by 3-4 elements. This is enough to retain high accuracy in the simulations.

Now, as was shown by Mathiesen et al. (see their Figure 2a), when using the above physical parameters, the average flow velocity, $\langle v \rangle$, satisfies
\begin{equation}\label{eq:averageVelocity}
    \langle v \rangle \lesssim  \frac{\gamma}{\mu_\text{w}} \frac{\text{Bo}}{100}.
\end{equation}
We therefore fix the velocity scale $v_0 \equiv \mu_\text{w}/(1000 \rho_\text{w} r)$ and choose the surface tension as $\gamma = 100 v_0 \mu_\text{w}/\text{Bo}$, for in that case the average flow velocity will be close to $v_0$, while the characteristic Reynolds number will be
\begin{equation}\label{eq:ReynoldsNumber}
    \text{Re} \equiv \frac{\rho_\text{w} r v_0}{\mu_\text{w}} \lesssim 10^{-3},
\end{equation}
ensuring that the flow is indeed creeping. We note that this choice implies that the Capillary number is $\text{Ca} = O(\text{Bo}/100)$, justifying the simplification of \eqref{eq:interfaceMomentumBalance} to \eqref{eq:interfacePressureDiff}. In addition, the velocity scale $v_0$ induces the advection time scale $t_0 \equiv 2 r / v_0$, and this allows us to define a stopping criterion for the simulations. When the simulation is started and the cluster is allowed to evolve from the circular configuration, its geometry initially undergoes rapid changes before approaching a fixed shape, in which the only changes in the phase field are due to diffusion at the cluster interface. Before the start of a simulation, we therefore draw a set $\{ \bm{r}_n \}$ of $1000 (L/r)^2$ uniformly distributed random points inside the domain and carry out the time integration up to $t = t_f$, where $t_f$ is the smallest time at which the phase field satisfies the criterion
\begin{equation}\label{eq:stoppingCriteria}
    \frac{1}{2} \max_{n} \Big( \big| \phi(\bm{r}_n, t_f) - \phi(\bm{r}_n, t_f - t_0) \big| \Big) < \frac{1}{100}.
\end{equation}
Finally, we choose the time step based on a CFL-type condition. To be precise, we take the time step to be $\Delta t = C_r h/v_0$, where $C_r \equiv 0.004/(r_\text{circ}/r)$ is the Courant number. Based on experience, this is approximately the largest possible time step allowed by the numerical time integration that ensures stability. We have chosen the Courant number to be inversely proportional to the cluster size in order to be able to simulate lower Bond numbers for larger clusters.  An example of a cluster's initial and final shapes are shown in the top row of Figure \ref{fig:phaseFieldStartAndEnd} for $r_\text{circ}/r = 4$, $\text{Bo} = 0.025$. The reader will notice that there appears to be slightly more wetting fluid in the final state than in the initial state. This is because the two fluids are not strictly incompressible in the phase field model. Instead, as a consequence of \eqref{eq:fullModel_mass} and \eqref{eq:fullModel_phi}, the phase field model conserves the total phase field in the sense that 
\begin{align}
    \frac{\diff }{\diff t} \int_\mathcal{D} \phi \, \diff^2 \bm{x} = 0,
\end{align}
in which $\mathcal{D}$ is the fluid domain. Therefore, the phase field model may lead to changes in cluster size. As discussed by \cite{YueEtAl2007}, this effect is particularly pronounced for large values of $\epsilon/r_\text{circ}$ and $L / r_\text{circ}$. In addition we have found that the surface-liquid interaction, enforced through \eqref{eq:fullModel_bc_phi}, also influences the final size of the cluster. With the presently used parameters the relative change in area is, however, small enough that we describe all cluster sizes in terms of their initial radii.

\subsection{Comparing the quasistatic model to the DNS model}
The purpose of the quasistatic model is to predict the critical Bond number at which a given cluster becomes unstable. The model will do so if it accurately predicts the evolution of the opening angles as a function of $\text{Bo}$, and we may therefore validate it by showing that its results for $\bm{\omega}(\text{Bo})$ are close to those of the DNS model. To do so, we first use the DNS model to compute the opening angles of two clusters for a discrete set of Bond numbers. One cluster starts out as a circle with $r_\text{circ}/r=2$ while the other is twice as large and has $r_\text{circ}/r = 4$. Since the interfaces in the DNS model are diffuse, we compute the opening angles from the intersection points between the obstacles and the curves along which $\phi = 0$. When measured in radians, the uncertainty of the opening angles is therefore approximately given by the width of the phase transition, $\epsilon$. Having solved the DNS model for a range of Bond numbers, we then initialize the quasistatic model from the opening angles produced by the DNS model for the smallest value of the Bond number and integrate \eqref{eq:simplifiedModel} with respect to $\text{Bo}$ using the classical fourth-order Runge-Kutta method with a step size of $\Delta \text{Bo} = 0.0001$. This value of $\Delta \text{Bo}$ is small enough to guarantee that the integration errors are negligible relative to the model errors.

\begin{figure}
    \centering
    \includegraphics[width=0.9\textwidth]{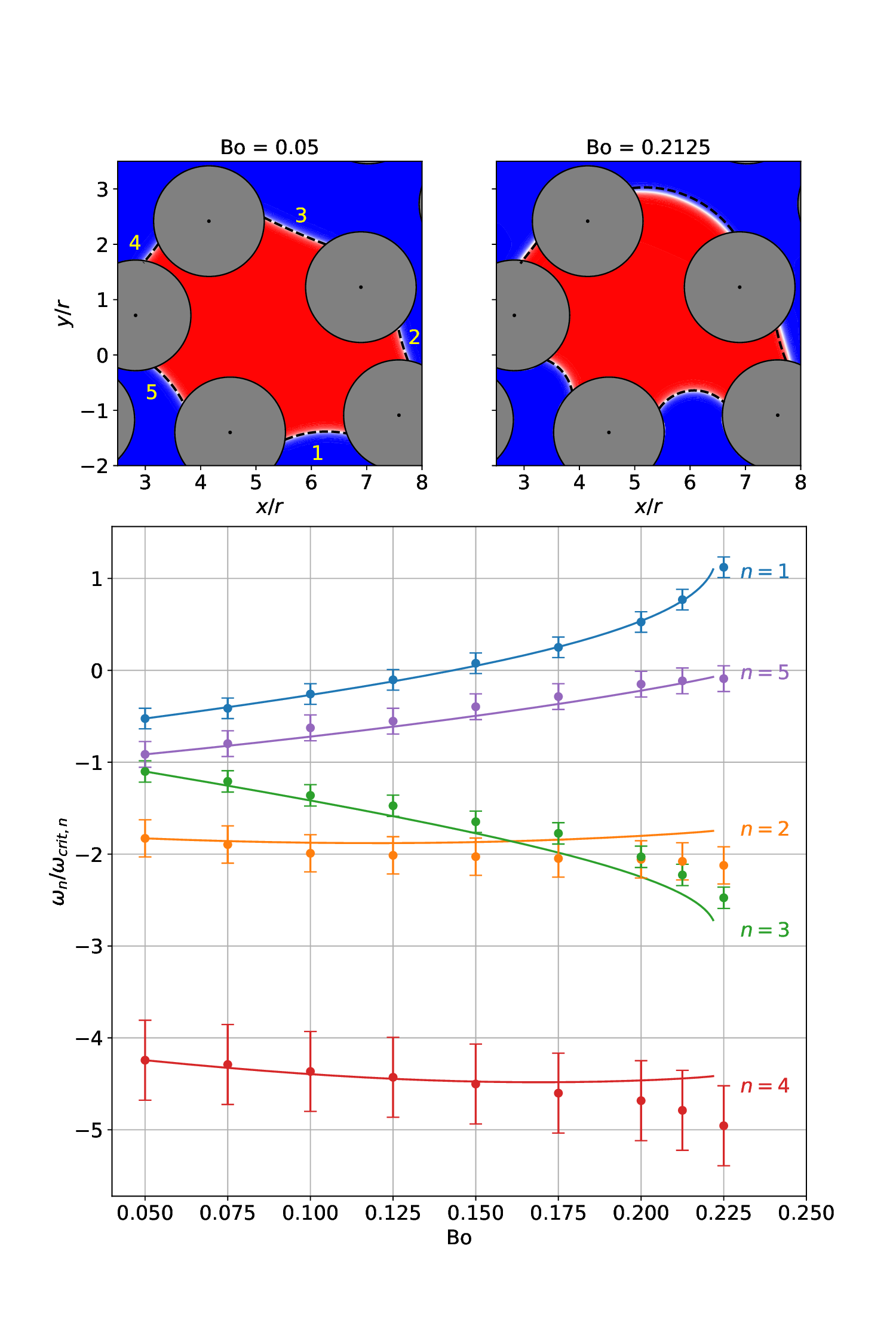}
    \caption{Top row: The phase field of a trapped cluster with $r_\text{circ}/r = 2$ for $\text{Bo} = 0.05$ (left) and $\text{Bo} = 0.2125$ (right). The dashed, black lines are the interfaces predicted by the quasistatic model. Since this model is initialized from the results of the DNS model for $\text{Bo} = 0.05$, the deviation between the models is zero at this Bond number. Bottom row: The evolution of the opening angles, $\omega_n$, normalized by their critical values given by \eqref{eq:omegaCrit} as a function of the Bond number. The full lines show the result of the quasistatic model, while the circles show the results of the DNS model. In the DNS computations, the uncertainty of $\omega_n$ is assumed to be $\pm \epsilon$ due to the diffusive interface, and this is indicated by the error bars. The numbering $n = 1, 2, ..., 5$ corresponds to the numbering of the cluster openings in the upper left figure.}
    \label{fig:validationrCluster2}
\end{figure}

The results for the cluster with $r_\text{circ}/r = 2$ are shown in Figure \ref{fig:validationrCluster2}. The agreement between the two models can be seen to be very good, with essentially only a few data points of the DNS model for interfaces 2 and 3 close to the critical Bond number failing to agree with the quasistatic model within the uncertainty. In particular, the quasistatic model's prediction of the evolution of $\omega_1$, which is the opening angle causing the cluster to become unstable, is very close to the prediction of the DNS model. While the quasistatic model predicts the critical Bond number of this cluster to be $\text{Bo}_\text{crit} = 0.222$, we have checked that the DNS model predicts instability for any Bond number larger than $0.2375$. As such, the quasistatic model's prediction of the critical Bond number deviates by maximally 6.5 \% from that of the DNS model, and we note that this deviation could be made smaller by carrying out additional simulations in the range $0.225 \le \text{Bo} \le 0.2375$.

\begin{figure}
    \centering
    \includegraphics[width=0.9\textwidth]{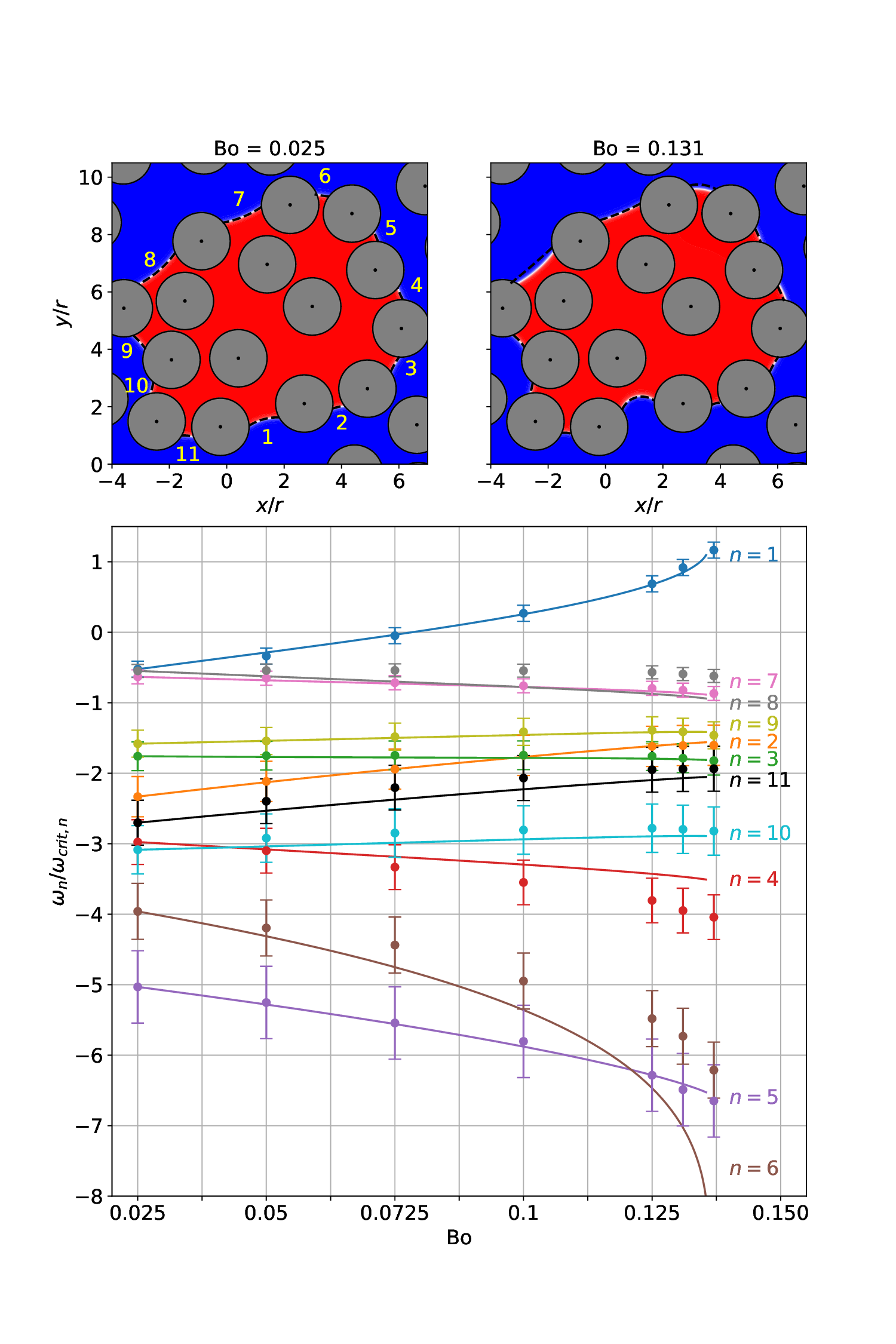}
    \caption{Top row: The phase field of a trapped cluster with $r_\text{circ}/r = 4$ for $\text{Bo} = 0.025$ (left) and $\text{Bo} = 0.131$ (right). The dashed, black lines are the interfaces predicted by the quasistatic model. Since this model is initialized from the results of the DNS model for $\text{Bo} = 0.025$, the deviation between the models is zero at this Bond number. Bottom row: The evolution of the opening angles, $\omega_n$, normalized by their critical values given by \eqref{eq:omegaCrit} as a function of the Bond number. The full lines show the result of the quasistatic model, while the circles show the results of the DNS model. In the DNS computations, the uncertainty of $\omega_n$ is assumed to be $\pm \epsilon$ due to the diffusive interface, and this is indicated by the error bars. The numbering $n = 1, 2, ..., 11$ corresponds to the numbering of the cluster openings in the upper left figure.}
    \label{fig:validationrCluster4}
\end{figure}

For the cluster with $r_\text{circ}/r = 4$ the story is very much the same, as can be seen in Figure \ref{fig:validationrCluster4}. In general, the quasistatic model and the DNS model agree quite closely, and the few existing deviations only appear as the Bond number approaches its critical value, this time mainly for interfaces 6 and 8. For this particular cluster, the quasistatic model predicts that the critical Bond number is $0.1356$, and since the DNS model yields instability for any Bond number larger than $0.143$, the deviation is maximally 4.5 \%. Again, this deviation could be reduced by executing the DNS model for more values of the Bond number.

This concludes our validation of the quasistatic model. Although more cases can always be checked, we consider the two presented here to provide sufficient evidence that the quasistatic model can in fact accurately predict the critical Bond number of a given cluster.

\section{The stability of a trapped cluster}\label{sec:stabilityProb}
We now turn our attention to the probability that a cluster of a given size is stable at a given Bond number. To compute this quantity, we integrate an ensemble of clusters to the point of breaking using the quasistatic model based on initial conditions produced by the DNS model. We find that the largest stable cluster size should scale in proportion to $r/\text{Bo}^{4/5}$, which is at odds with both MFA1 and MFA2 described in Section \ref{sec:introduction}. For that reason, we elaborate on the reasons why these arguments do not lead to the correct result. We end this section by showing that the $r/\text{Bo}^{4/5}$-scaling is not a general result; indeed, we compute its dependency on the ratio between the densities of the two fluids, $\rho_\text{w}/\rho_\text{nw}$, explicitly and argue that the exponent is not unique but a function of the system's parameters.

\subsection{The probability that a cluster is stable as a function of its size and the Bond number}
In order to compute the probability that a cluster of a given size is stable for a given Bond number, we first construct initial conditions for the quasistatic model using the DNS model. For each of the cluster sizes $r_\text{circ}/r = 2$, 4, 6, 8 and 10 we employ the DNS model with $\text{Bo} = (10 r_\text{circ}/r)^{-1}$ to compute the stable configuration of 10 clusters starting at random positions in the domain. Each stable state is then used to initialize the quasistatic model, which is subsequently integrated up to its critical Bond number, $\text{Bo}_\text{crit}$. If we let $N_\text{stable}(r_\text{circ}/r, \text{Bo})$ denote the number of clusters of size $r_\text{circ}$ which have $\text{Bo}_\text{crit} > \text{Bo}$, this enables us to compute the probability, $P$, that a cluster of size $r_\text{circ}/r$ is stable at Bond number $\text{Bo}$ as 
\begin{equation}\label{eq:stabilityProbability}
    P(r_\text{circ}/r, \text{Bo}) 
    = \frac{N_\text{stable}(r_\text{circ/r}, \text{Bo})}{10}.
\end{equation}
The choice of using only 10 trapped clusters for each size is a consequence of our available computing resources; using four Intel Xeon Gold 6148 CPUs having 20 cores each, it took us approximately four weeks to generate the initial conditions for the case $r_\text{circ}/r = 10$ alone. Generating, for example, 100 initial conditions would hence be unfeasible without significantly larger computational resources. In contrast, the part of the computation executed using the quasistatic model takes less than a minute on a single CPU core even for the largest cluster size, the ``bottleneck'' being the extraction of the opening angles from the DNS model. Hence, the computation of \eqref{eq:stabilityProbability} would have been unfeasible even for a modest resolution in $\text{Bo}$ without the quasistatic model.

\begin{figure}
    \centering
    \includegraphics[width=0.72\textwidth]{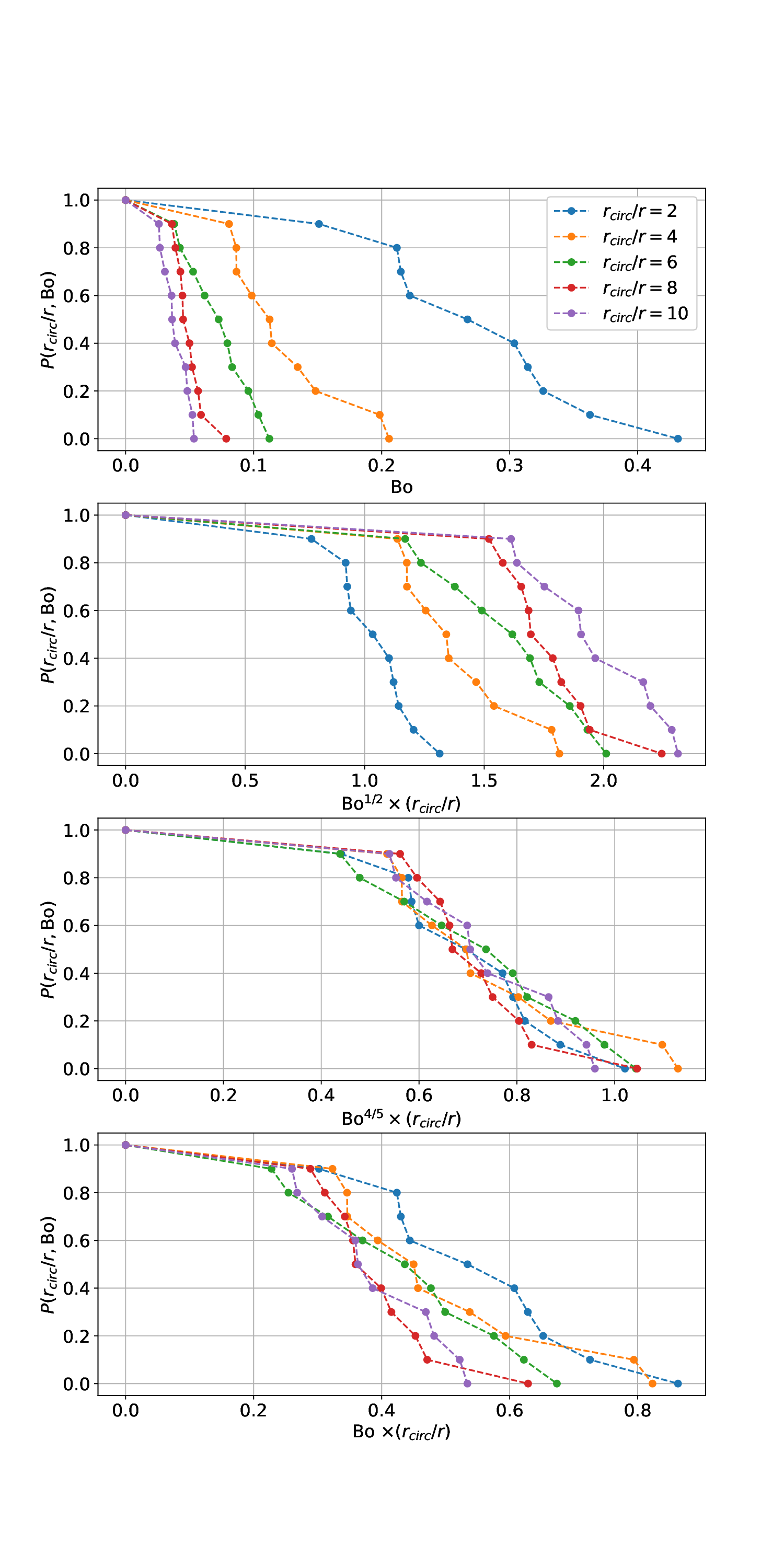}
    \caption{The probability that a cluster of size $r_\text{circ}/r$ is stable at a given Bond number as a function of (from top to bottom) $\text{Bo}$, $\text{Bo}^{1/2} \times (r_\text{circ}/r)$, $\text{Bo}^{4/5} \times (r_\text{circ}/r)$ and $\text{Bo} \times (r_\text{circ}/r)$. The legend applies to all figures. According to MFA1 and MFA2, the results for $P(r_\text{circ}/r, \text{Bo)}$ should collapse to a single curve when plotted as a function of $\text{Bo} \times (r_\text{circ}/r)$ and $\text{Bo}^{1/2} \times (r_\text{circ}/r)$, respectively.}
    \label{fig:stabilityProbability}
\end{figure}

Now, our results for the probability of stability are shown in Figure \ref{fig:stabilityProbability}. When the Bond number is fixed, the uppermost subplot of this figure illustrates the fact that large clusters are more prone to be unstable than small clusters, as one would expect. However, our results do not agree with the predictions of MFA1 and MFA2 described in Section \ref{sec:introduction}. According to MFA2 the probability \eqref{eq:stabilityProbability} should be a function of $\text{Bo}^{1/2} \times (r_\text{circ}/r)$ only, and the results for $P(r_\text{circ}/r, \text{Bo)}$ should therefore collapse to a single curve when plotted against this parameter. As is clearly seen in the second subplot of Figure \ref{fig:stabilityProbability}, our results do not have this property. Likewise, as shown in the lowermost subplot of the figure, our results also do not follow the scaling predicted by MFA1, i.e. that the probability should be a function of the parameter $\text{Bo} \times (r_\text{circ}/r)$ only. The disagreement with MFA2 is, of course, particularly noteworthy, since we have used the same fluid properties as \cite{MathiesenEtAl2023}, and because they found a very good agreement between MFA2 and their numerical simulations. In that connection, it is, however, important to note that Mathiesen et al. simulated a dynamic flow in a very large domain where clusters could both split and merge continuously. Moreover, their domains contained a large number of both trapped and mobilized clusters, which they did not distinguish between when comparing to the analytical predictions. As such, the agreement found by \cite{MathiesenEtAl2023} may be induced by their dynamic setting.

By varying the exponent of $\text{Bo}$, we find that the best collapse of our data to a single curve is found when it is plotted as a function of $\text{Bo}^{4/5} \times (r_\text{circ}/r)$. As shown in the second-lowest subplot of Figure \ref{fig:stabilityProbability}, the collapse is reasonably good for this exponent of $\text{Bo}$, and it means that the size of the largest stable cluster should scale in proportion to $r/\text{Bo}^{4/5}$. However, this result only holds for the set of fluid properties that are presently used.

\subsection{Shortcomings of the mean-field theories}
Having shown that both MFA1 and MFA2 do not agree with the simulated size of the largest stable cluster, we now discuss some reasons for why this is the case.

\subsubsection{The pressure field around a trapped cluster}
In any mean-field theory, the pressure field surrounding the cluster is approximated through the average pressure gradient of the flow. For the present gravity-driven setting this corresponds to the assumption that the difference between the pressure just outside of the $n$th and $i$th openings of the cluster is given by
\begin{equation}\label{eq:averagePessureDifference}
    p_{\text{out},n} - p_{\text{out},i} = \rho_\text{nw} g (y_{\text{out},n} - y_{\text{out},i}).
\end{equation}
Unless the cluster is large, this equation will only provide a poor approximation for the pressure difference between its top and bottom. In essence the explanation for this is that the pressure difference between two different openings is a random variable, whose standard deviation is independent of the difference between the two openings' $y$-coordinates. While \eqref{eq:averagePessureDifference} can in fact provide a reasonably good estimate of the pressure difference's mean value, the mean-field approximation can only be considered accurate if the standard deviation is small relative to the mean value. Since this requires a large difference in $y$-coordinates, the mean-field approximation can therefore only be accurate for large clusters.

\begin{figure}
    \centering
    \includegraphics[width=0.9\textwidth]{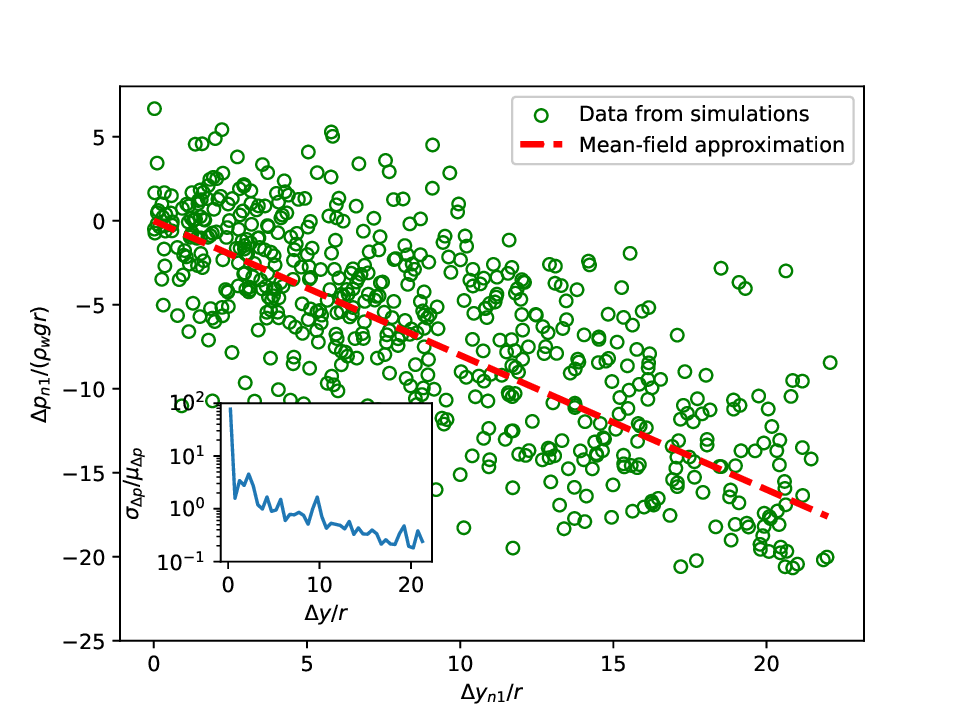}
    \caption{Main figure: the pressure difference $\Delta p_{n1} \equiv p_{\text{out},n} - p_{\text{out},1}$ as a function of the difference in $y$-coordinates between the evaluation points, $\Delta y_{n1} \equiv y_{\text{out},n} - y_{\text{out},1}$. The green circles are obtained from the initial conditions produced by the DNS model using the outside evaluation point $\bm{r}_{\text{out},n} \equiv \bm{r}_{\text{int},n} - \delta \bm{\hat{n}}_n$ with $\delta = 3 \epsilon$. The red, dashed line shows the mean-field approximation given by \eqref{eq:averagePessureDifference}. Inset: the standard deviation, $\sigma_{\Delta p}$, of the green circles divided by the mean value, $\mu_{\Delta p}$, of the pressure difference as a function of the difference in $y$-coordinates. These quantities are computed from the green circles for each of the segments $m/2 \le \Delta y/r < (m+1)/2$ for $m = 0, 1, 2, ...$. Roughly we have $\sigma_{\Delta p} \simeq 3 \rho_\text{w} g r $ independently of $\Delta y$.}
    \label{fig:isTheAveragePressureGradientAGoodApproximation}
\end{figure}

To illustrate this point for the presently used porous geometry, Figure \ref{fig:isTheAveragePressureGradientAGoodApproximation} shows the pressure differences between the openings obtained from the initial conditions produced by the DNS model used for Figure \ref{fig:stabilityProbability} as a function of their difference in $y$-coordinate. From this it is clear that the pressure difference is a random variable whose mean value is reasonably well described by \eqref{eq:averagePessureDifference} while its standard deviation is roughly $3 \rho_\text{w}gr$ independently of the difference in $y$-coordinates. This means that if we want the standard deviation to be smaller than, say, $5\%$ of the mean-field prediction, we must have $(y_{\text{out},n} - y_{\text{out},1})/r \simeq 75$. Thus the pressure difference between the top and bottom of a particular cluster can only be expected to be within $5\%$ of the mean-field prediction if the cluster's size is $r_\text{circ}/r \gtrsim 32$, which is indeed quite large. The inset in Figure \ref{fig:isTheAveragePressureGradientAGoodApproximation} shows the ratio between the standard deviation and the mean value of the pressure difference as a function of the difference in $y$-coordinates. Clearly this ratio diverges for small differences in $y$-coordinates, and for none of the clusters considered in this study is the ratio smaller than $10\%$.

\begin{figure}
    \centering
    \includegraphics[width=0.9\textwidth]{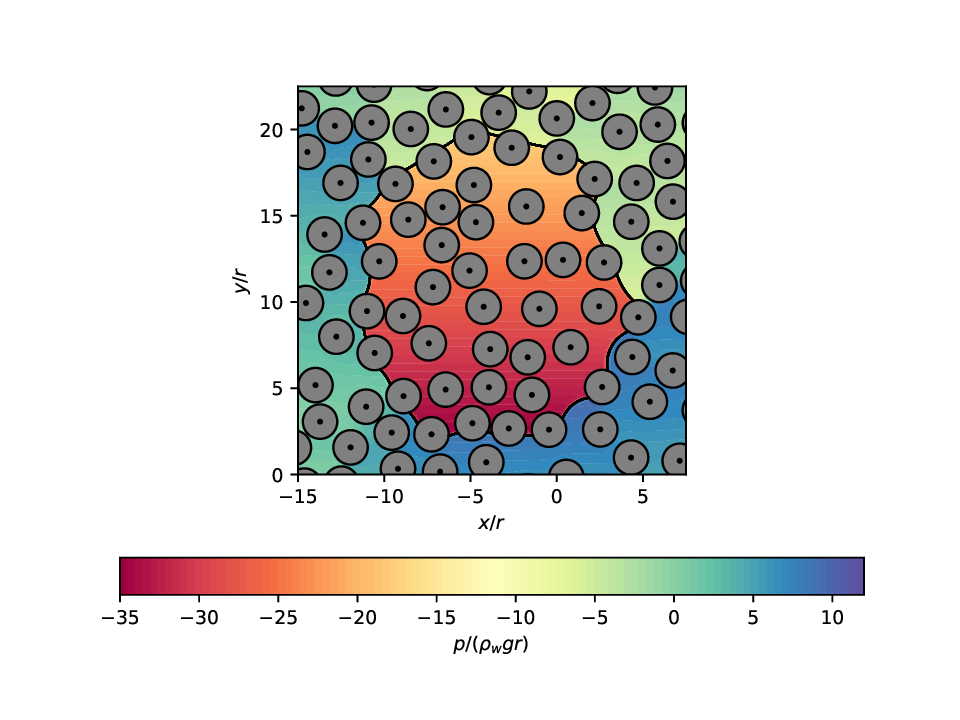}
    \caption{The pressure field around a trapped cluster of size $r_\text{circ}/r = 8$.}
    \label{fig:pressureFieldTrappedCluster}
\end{figure}

The physical reason for the fixed standard deviation of the pressure difference is that locally the pressure field is much more affected by the geometry of the medium than by the average pressure gradient of the flow. This is illustrated by the trapped cluster depicted in Figure \ref{fig:pressureFieldTrappedCluster}, for which the mean-field assumption is clearly inaccurate. Instead, the pressure field remains approximately constant over, say, 5-10 cluster openings before changing abruptly because the surrounding fluid must go through a narrow passage. For other, more permeable porous geometries there will be fewer of these abrupt changes, and the standard deviation of the pressure differences between the cluster openings is therefore expected to be smaller. As a consequence, the mean-field approximation \eqref{eq:averagePessureDifference} is expected to become more accurate the more permeable the porous media is. For example, it may work well for all but the smallest clusters in the highly permeable media used by \cite{TallakstadEtAl2009}.

\subsubsection{The width of the breaking opening}
\begin{figure}
    \centering
    \includegraphics[width=0.72\textwidth]{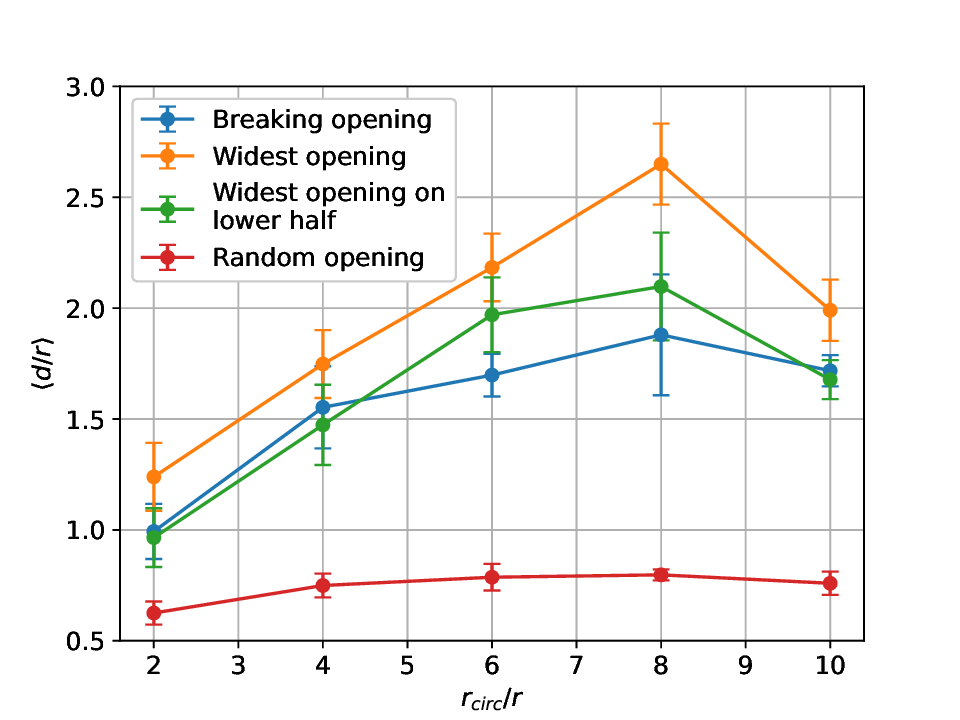}
    \caption{The average width of the breaking opening, the widest opening, the widest opening on the lower half of the cluster as well as a random opening as functions of the cluster size $r_\text{circ}/r$. The uncertainty indicated by the error bars is the estimated standard deviation divided by the square root of the number of samples.}
    \label{fig:openingWidths}
\end{figure}

In MFA1 it is assumed that a cluster breaks when the pressure difference (approximated through the average pressure gradient) across the cluster is larger than the maximum sustainable pressure difference of a typical opening. If this was true, one would expect that the width of the opening that actually breaks on average is the same as the average width of a random cluster opening. As illustrated in Figure \ref{fig:openingWidths}, the average width of the breaking opening increases from $\langle d_\text{break}/r \rangle \simeq 1.0$ for small clusters to $\langle d_\text{break}/r \rangle \simeq 1.7$ for large clusters. In contrast, the average width of a random opening is nearly independent of the cluster size and has $\langle d_\text{random}/r \rangle \simeq 0.8$. As such, this assumption made in MFA1 is not generally accurate.

Interestingly, it is assumed in MFA2 that a cluster breaks if its widest opening cannot sustain the pressure difference across the cluster. Considering the results shown in Figure \ref{fig:openingWidths}, the width of the breaking opening is on average clearly smaller than the average width of the widest opening. However, the breaking opening is most often located on the lower half of the cluster, and the average width of the widest opening on the lower half of the cluster is typically quite close to the average width of the breaking opening. In that sense, the assumption made in MFA2 can be said to be satisfied.

\subsubsection{The probability density function of the maximum sustainable pressure difference of a random opening}
\begin{figure}
    \centering
    \includegraphics[width=0.72\textwidth]{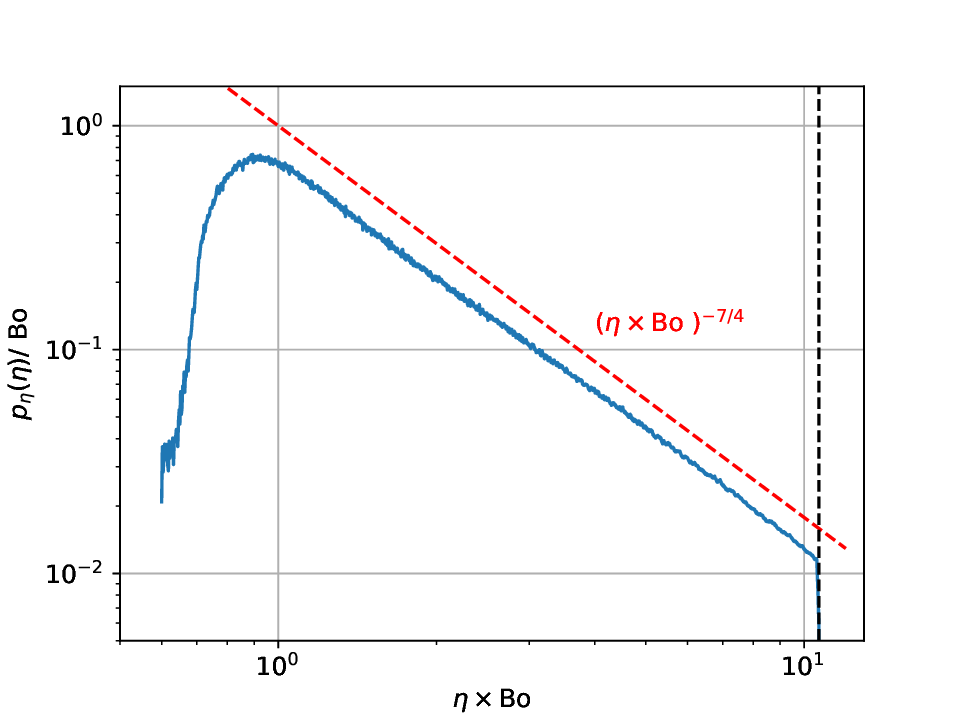}
    \caption{The probability density function of the non-dimensional maximum sustainable pressure, $\eta$, for a random cluster opening given by \eqref{eq:etaPdf} for the contact angle $\theta_0 = \pi/3$. The PDF of $d/r$ in this expression is sampled from the edge lengths of a Delaunay triangulation of the porous geometry, and the uncertainty induced by the finite number of edges is what causes the small, local oscillations of the curve. The cutoff indicated by the dashed, black line is a consequence of the requirement that the distance between any two obstacles, $d$, must be greater than the minimum distance $d_\text{min} = r/10$. The dashed, red line is the function $(\eta \times \text{Bo})^{-7/4}$ and shows that $p_\eta$ asymptotically (but prior to the cutoff) follows a power law.}
    \label{fig:etaPDF}
\end{figure}

In Section \ref{sec:introduction}, we remarked that the exponential distribution for the maximum sustainable pressure of a random opening assumed in MFA2 is not supported by theory. Here we go one step further and argue that the assumption is in fact incorrect. To that end, we start by defining the maximum non-dimensional sustainable pressure difference $\eta \equiv \text{max} (\Delta p /(\rho_\text{w} g r))$. Inserting \eqref{eq:omegaCrit} into \eqref{eq:youngLaplace2} leads to
\begin{equation}\label{eq:maximumPressureDiff}
    \eta = \frac{2}{\text{Bo}} \frac{1}{\big( (d/r + 2)^2 - 4 \sin^2(\theta_0) \big)^{1/2} - 2\cos(\theta_0)}
\end{equation}
for an opening of width $d$. Since the opening width is a random variable, $\eta$ is likewise a random variable, and one can show that its PDF is given by 
\begin{equation}\label{eq:etaPdf}
    p_\eta(\eta) = 
    \frac{\text{Bo} \big( 2 + 2\cos(\theta_0) \eta \text{Bo} \big)}{(\eta \text{Bo})^2 \big( 1 + 2\cos(\theta_0) \eta \text{Bo} + (\eta \text{Bo})^2\big)^{1/2}}
    \times 
    p_{d/r} \Big( \frac{2}{\eta \text{Bo}} \big( 1 + 2\cos(\theta_0) \eta \text{Bo} + (\eta \text{Bo})^2 \big)^{1/2} - 2 \Big),
\end{equation}
in which $p_{d/r}$ is the PDF of $d/r$. To the best of our knowledge $p_{d/r}$ is not known analytically, and the amount of data that we have available for each value of $r_\text{circ}/r$ does not allow us to sample it accurately. We therefore resort to an approximation and use the PDF of the edge length (minus $2r$) from Delaunay triangulations of the obstacle centers of a large number of porous geometries. In this connection one should note that $p_{d/r}$ is expected to be smaller for small values of $d/r$ when based on the Delaunay triangulation than on actual trapped clusters. This is so because a trapped cluster tries to minimize its surface length in order to reach a local energy minimum and because the Delaunay algorithm tries to avoid short edges in order to maximize the minimum angle of the triangulation. As a consequence, $p_\eta$ is expected to have a heavier tail (that is, to decay less rapidly) when based on actual trapped cluster openings than on Delaunay edges. Our result for the PDF of $\eta$ is shown in Figure \ref{fig:etaPDF} for the contact angle $\theta_0 = \pi/3$. From the figure it is clear that $p_\eta$ falls off as $\text{Bo}^{-3/4} \times \eta^{-7/4}$ asymptotically when based on Delaunay edges, and when combined with the above remark, it is safe to conclude that $p_\eta$ does not fall off exponentially as a function of $\eta$ if it (more accurately) had been computed from actual trapped cluster openings. As such, the assumption in MFA2 made about the form of $p_\eta$ cannot be correct. In this connection it is interesting to note that if MFA2 is carried out with a distribution of the form $p_\eta(\eta) \propto \text{Bo}^{\beta+1} \times \eta^{\beta}$ (which is implied by \eqref{eq:etaPdf} for large values of $\text{Bo} \times \eta$ if $p_{d/r}$ follows a power law distribution), one finds that the largest stable cluster size scales is very well approximated by $r/\text{Bo}$ for the relevant range of Bond numbers regardless of the value of $\beta$, as long as $\beta < -1$. Hence, if the maximum sustainable pressure difference of a random opening follows a power law distribution, MFA2 in fact gives the same result as MFA1.

\subsection{The $r/\text{Bo}^{4/5}$-scaling of the largest stable cluster size is not a general result}
The value $4/5$ for the exponent of $\text{Bo}$ found to give the best data collapse in Figure \ref{fig:stabilityProbability} is implied by the particular values of the fluid properties. It is not a general result valid over a large part of the system's parameter space.

\begin{figure}
    \centering
    \includegraphics[width=0.72\textwidth]{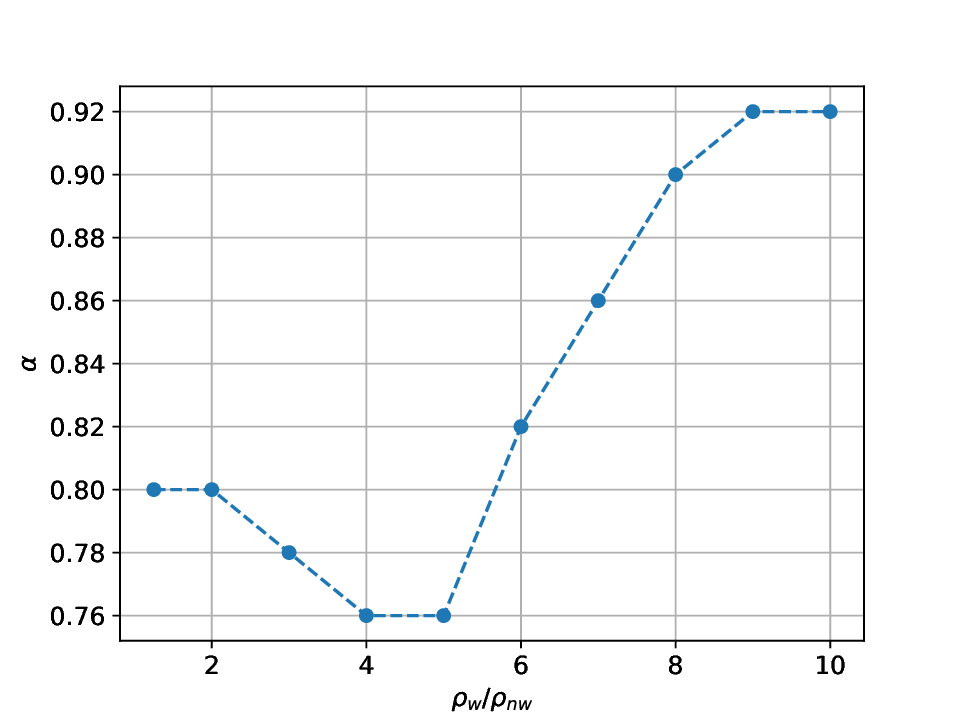}
    \caption{As a function of the density ratio, $\rho_\text{w}/\rho_\text{nw}$, the figure shows the value of $\alpha$ that leads to the best possible data collapse of the probability that a cluster of size $r_\text{circ}/r$ is stable when plotted as a function $\text{Bo}^\alpha \times (r_\text{circ}/r)$. }
    \label{fig:alphaVwRhoRatio}
\end{figure}

To back up this claim, we consider the exponent's dependency on the density ratio. To compute the critical Bond number for an arbitrary density ratio we have first integrated the states computed by the DNS model for $\rho_\text{w}/\rho_\text{nw} = 1.25$ used to produce Figure \ref{fig:stabilityProbability} up to different values of $\rho_\text{w}/\rho_\text{nw}$ using \eqref{eq:simplifiedModelRho} before integrating the resulting states up to the critical Bond numbers using \eqref{eq:simplifiedModel}. Our results for the exponent, $\alpha$, which gives the best collapse of the probability that a cluster of size $r_\text{circ}/r$ is stable when plotted against the parameter $\text{Bo}^{\alpha} \times (r_\text{circ}/r)$, are shown in Figure \ref{fig:alphaVwRhoRatio} in the range $1.25 \le \rho_\text{w}/\rho_\text{nw} \le 10.0$. From this it is evident that the exponent is a nontrivial function of the density ratio. While we do not know the reason for the local minimum of $\alpha$ around $\rho_\text{w}/\rho_\text{nw} \simeq 5$, it is relatively simple to understand why $\alpha$ approaches a value close to $1$ as the density ratio becomes large. In the limit of large density ratio, $p_{\text{in},n} - p_{\text{in},i}$ is much larger in magnitude than $p_{\text{out},n} - p_{\text{out},i}$ for any two openings $n$ and $i$. As an approximation, one may therefore treat the pressure field outside the cluster as being constant in this limit. This means that if two openings have the same width, the one of the two located the farthest upstream will break first. For that reason it is somewhat reasonable to assume that, among all openings, it is the one lying the most upstream that breaks first. Since the width of the most upstream opening is distributed just like any other opening, it also has the same average properties, and the limit of large density ratio therefore approximately corresponds to the case analyzed in MFA1. One therefore ends up with $\alpha$ being close to $1$ when the density ratio is large.

In this connection it is interesting to note, that \cite{TallakstadEtAl2009} experimentally verified that the predictions of MFA1 hold for $\rho_\text{w}/\rho_\text{nw} \simeq 10^{-3}$ in a highly permeable medium. The explanation for this is as follows: when the trapped fluid is much lighter than the surrounding fluid, the internal pressure field becomes negligible. Moreover, if the medium is highly permeable, the surrounding pressure field is well approximated by the average pressure gradient of the flow. As a consequence, if two openings have the same width, the one of the two located the farthest upstream will break first. The argument used above then implies that $\alpha$ will be close to $1$.

Similar quantitative analyses as the one above could, of course, be carried out for $\alpha$'s dependency on other parameters such as the contact angle, $\theta_0$, the minimum allowed distance between two obstacles, $d_\text{min}$, and other geometrical variables. While we have not done so, it is seems unlikely that $\alpha$ should be independent of these variables. For example, as we have already argued, the minimum distance is a key parameter when it comes to the adequacy of pressure-field description in the mean-field arguments. When this distance is large, the optimal exponent will likely approach the prediction of the mean-field theory, while it remains unknown what happens in the limit where this distance is small. Classifying cluster stability in terms of a scaling relation of the form $r/\text{Bo}^{\alpha}$ therefore requires that $\alpha$ is allowed to depend on several variables, leaving the efficiency of this representation questionable.

\section{Conclusions and outlook}\label{sec:conclusions}
In this paper, we have investigated the stability of trapped clusters in two-phase flow in porous media. We have introduced the quasistatic model \eqref{eq:simplifiedModel} that captures how cluster geometry evolves with changes in the Bond number. Given a specific cluster configuration at a known Bond number, our model predicts the critical Bond number at which an opening in the cluster becomes unstable and is pushed inward. This approach significantly reduces computational cost compared to full hydrodynamic simulations, in particular when extended to other parameters such as done in \eqref{eq:simplifiedModelRho} for the density ratio. As such, the model offers a practical avenue for analyzing cluster stability. 

Using our quasistatic model in combination with initial conditions generated using phase-field modelling, we have computed the probability that a cluster is stable as a function of the Bond number for different values of its size $r_\text{circ}$. We have found that these curves collapse to a single one when plotted against the parameter $\text{Bo}^{4/5} \times (r_\text{circ}/r)$; when plotted against $\text{Bo} \times (r_\text{circ}/r)$ and $\text{Bo}^{1/2} \times (r_\text{circ}/r)$, as predicted by the mean-field arguments MFA1 and MFA2, respectively, there is no collapse. The exponent $4/5$ is, however, not a general result valid over a large subset of the system's parameter space; instead, it is a result of the chosen parameters for the fluids and the geometry. Using \eqref{eq:simplifiedModelRho} to vary  the ratio between the densities of the two fluids, we have shown that the exponent is a non-trivial function of this ratio, and we have noted that the exponent likely also depends on several other variables. The representation of the largest stable cluster size in terms of a scaling relation of the form $r/\text{Bo}^{\alpha}$ is therefore questionable.

As a final remark, we note that the quasistatic model can be generalized to a three-dimensional setting. For realistic porous media in 3D, we expect, however, that the analytical approach applied in this work will not be entirely sufficient, since such geometries are, as a starting point, represented by a voxel grid. Instead, it is likely that the geometry of the cluster must be represented in some discrete way, for example through a set of grid points and their connectivity. The change in representation should not affect the computational efficiency of the model. As in two dimensions, the success of the model in 3D will hinge on the assumption that the left hand side of \eqref{eq:pressureCurvatureRelation} is independent of the Bond number. In two dimensions this assumption holds because no pathways are created nor destroyed before the point of breaking. In three dimensions it, however, remains to be seen whether this is enough to keep the normalized pressure difference between two cluster openings (approximately) constant in the range $0 \le \text{Bo} \le \text{Bo}_\text{crit}$. After all, three-dimensional media are known to be much more connected than their two-dimensional counterparts. 
\\
\\

\noindent
{\small \textbf{Funding.} M.K. and J.M. acknowledge support from the ESS lighthouse on hard materials in 3D, SOLID, funded by the Danish Agency for Science and Higher Education, Grant 8144-00002B. G.L. acknowledges support from the Research Council of Norway through the PoreLab Center of Excellence (grant number 262644) and the Researcher Project for Young Talents M4 (grant number 325819) and the Norwegian Centre of Advanced Study through the Young CAS project \textit{Mixing by interfaces}. T.L.B. acknowledges support from the Research Council of Norway through the project MinMix (grant number 353372).}\\

\noindent
{\small \textbf{Declaration of Interests.} The authors report no conflict of interest.}


\appendix 
\section{A set of equations predicting the change in cluster geometry as a function of the density ratio}\label{app:simplifiedModelRho}
When varying the density ratio, $\rho_\text{w}/\rho_\text{nw}$ (by keeping $\rho_\text{w}$ fixed and varying $\rho_\text{nw}$), while all other parameters are held constant, the change in a trapped cluster's geometry can be predicted by solving a set of equations analogous to \eqref{eq:simplifiedModel}. As the mass and therefore also the area of the cluster is independent of this ratio, it is straightforward to show that 
\begin{equation}\label{eq:areaIndependentOfRhoRatio}
        0 = \sum_{n = 1}^{N} \bigg( \frac{\partial f_n}{\partial \omega_n} + \frac{\partial f_{n+1}}{\partial \omega_n} \bigg) \frac{\partial \omega_n}{\partial (\frac{\rho_\text{w}}{\rho_\text{nw}})},
\end{equation}
corresponding to \eqref{eq:clusterArea3}. Setting $i = 1$ in \eqref{eq:pressureCurvatureRelation} and taking the partial derivative with respect to $\rho_\text{w}/\rho_\text{nw}$ gives that
\begin{equation}\label{eq:pressureConservationRho}
\begin{aligned}
    \bigg( \frac{\diff (y_{\text{int},1}/r)}{\diff \omega_1} & + \frac{1}{\text{Bo}} \frac{\diff (\kappa_1 r)}{\diff \omega_1} \bigg) \frac{\partial \omega_1}{\partial (\frac{\rho_\text{w}}{\rho_\text{nw}})}
    - 
    \bigg( \frac{\diff (y_{\text{int},n}/r)}{\diff \omega_n} + \frac{1}{\text{Bo}} \frac{\diff (\kappa_n r)}{\diff \omega_n} \bigg) \frac{\partial \omega_n}{\partial (\frac{\rho_\text{w}}{\rho_\text{nw}})}
    \\
    & = 
    - \Big( \frac{\rho_\text{w}}{\rho_\text{nw}} \Big)^{-1} \frac{p_{\text{out},1} - p_{\text{out},n}}{\rho_\text{w} g r},
\end{aligned}
\end{equation}
for $n = 2, 3, ..., N$. In combination, \eqref{eq:areaIndependentOfRhoRatio} and \eqref{eq:pressureConservationRho} constitute the system of $N$ coupled equations \eqref{eq:simplifiedModelRho} governing the evolution of the opening angles as a function of the density ratio. As is evident from \eqref{eq:areaIndependentOfRhoRatio} and \eqref{eq:pressureConservationRho}, one ends with the same matrix, $\mathsfbi{A}$, as in \eqref{eq:simplifiedModel} when writing these equations in matrix form; the right hand side vector is, however, different. Defining 
\begin{equation}\label{eq:cRho}
    \bm{c}_{\rho_\text{w}/\rho_\text{nw}} = 
    \begin{bmatrix}
        0   \\
        c_2^{(\rho_\text{w}/\rho_\text{nw})} \\
        c_3^{(\rho_\text{w}/\rho_\text{nw})} \\
        \vdots \\
        c_{N-1}^{(\rho_\text{w}/\rho_\text{nw})} \\
        c_{N}^{(\rho_\text{w}/\rho_\text{nw})}
    \end{bmatrix}
    , 
    \qquad 
    \text{where}
    \qquad
    c_n^{(\rho_\text{w}/\rho_\text{nw})} = - \Big( \frac{\rho_\text{{w}}}{\rho_\text{nw}} \Big)^{-1} \frac{p_{\text{out},1} - p_{\text{out},n}}{\rho_\text{w} g r}
\end{equation}
gives the matrix form of the equations stated in \eqref{eq:simplifiedModelRho}.

\begin{figure}
    \centering
    \includegraphics[width=0.9\textwidth]{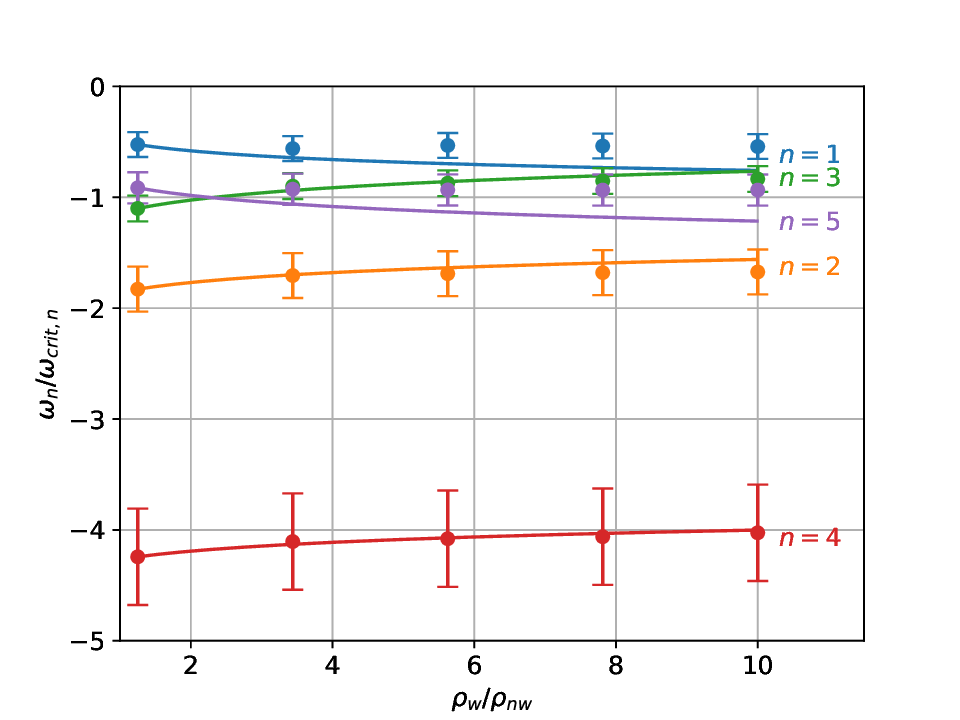}
    \caption{The evolution of the opening angles, $\omega_n$, normalized by their critical values given by \eqref{eq:omegaCrit} as a function of the density ratio for the same trapped cluster as in Figure \ref{fig:validationrCluster2} with $\text{Bo} = 0.05$. The full lines show the result of the quasistatic model \eqref{eq:simplifiedModelRho}, while the circles show the results of the DNS model. In the DNS computations, the uncertainty of $\omega_n$ is assumed to be $\pm \epsilon$ due to the diffusive interface, and this is indicated by the error bars. The numbering $n = 1, 2, ..., 5$ corresponds to the numbering of the cluster openings applied in Figure \ref{fig:validationrCluster2}.}
    \label{fig:validationRho}
\end{figure}

To validate that these equations indeed predict the change of shape of a cluster, we consider the same cluster as in Figure \ref{fig:validationrCluster2} with $\text{Bo} = 0.05$ and compare its geometry to the predictions of the DNS model as the density ratio is increased from $\rho_\text{w}/\rho_\text{nw} = 1.25$ to $\rho_\text{w}/\rho_\text{nw} = 10.0$. When calculating the entries of the vector \eqref{eq:cRho} we evaluate the pressure at $\bm{r}_{\text{out},n} \equiv \bm{r}_{\text{int},n} - \delta \bm{\hat{n}}_n$ with $\delta = 3\epsilon$, corresponding to the evaluation point being three interface thicknesses (or, measured in a different way, 15\% of an obstacle radius) away from the interface. The result of this computation is shown in Figure \ref{fig:validationRho}, which illustrates the fact that \eqref{eq:simplifiedModelRho} can indeed be used to accurately estimate the geometrical change over a relatively large range of density ratios.

\begin{thebibliography}{40}

\bibitem[Abels et al.\ (2012)]{AbelsEtAl2012}
{\sc Abels, H., Garcke, H. \& Grün, G.} 2012 {Thermodynamically Consistent, Frame Indifferent Diffuse Interface Models for Incompressible Two-Phase Flows with Different Densities}, {\it Math. Mod. and Meth. in Apl. Sci.} {\bf 22}, 1150013.


\bibitem[Alamooti et al.\ (2023)]{AlamootiEtAl2023}
{\sc Alamooti, A., Colombano, S., Glabe, Z. A., Lion, F., Davarzani, D., \& Ahmadi-Sénichault, A.} 2023 {Remediation of multilayer soils contaminated by heavy chlorinated solvents using biopolymer-surfactant mixtures: two-dimensional flow experiments and simulations}, {\it Water Research}, {\bf 243}, 120305.

\bibitem[Aland \& Voigt (2012)]{AlandVoigt2012}
{\sc Aland, S., \& Voigt, A.} 2012 {Benchmark computations of diffuse interface models for two‐dimensional bubble dynamics}, {\it Int. J. Num. Meth. Fluids}, {\bf 69}(3), 747-761.


\bibitem[Avraam \& Payatakes (1995)]{AvraamPayatakes1995}
{\sc Avraam, D. G. \& Payatakes, A. C.} 1995 {Flow regimes and relative permeabilities during steady-state two-phase flow in porous media}, {\it J. Fluid Mech.} {\bf 293}, 207-236.



\bibitem[Brakke\ (1992)]{Brakke1992}
{\sc Brakke, K. A.} 1992 {The surface evolver}, {\it Exp. Math.} {\bf 1}(2), 141-165.


\bibitem[Carlson et al.\ (2012)]{CarlsonEtAl2012}
{\sc Carlson, A., Bellani, G., \& Amberg, G.} 2012 {Universality in dynamic wetting dominated by contact-line friction}, {\it Phys. Rev. E}, {\bf 85}(4), 045302.


\bibitem[Cieplak \& Robbins\ (1988)]{CieplakRobbins1988}
{\sc Cieplak, M. \& Robbins, M. O.} 1988 {Dynamical Transition in Quasistatic Fluid Invasion in Porous Media}, {\it Phys. Rev. Lett.} {\bf 60}, 2042-2046.


\bibitem[Cieplak \& Robbins\ (1990)]{CieplakRobbins1990}
{\sc Cieplak, M. \& Robbins, M. O.} 1990 {Influence of contact angle on quasistatic fluid invasion of porous media}, {\it Phys. Rev. B} {\bf 41}, 11508-11522.


\bibitem[Ding et al.\ (2007)]{DingEtAl2007}
{\sc Ding, H., Spelt, P. D., \& Shu, C.} 2007 {Diffuse interface model for incompressible two-phase flows with large density ratios}, {\it J. Comput. Phys.} {\bf 226}(2), 2078-2095.


\bibitem[Feder\ (1980)]{Feder1980}
{\sc Feder, J.} 1980 {Random sequential adsorption}, {\it J. Theor. Biol.}, {\bf 87}(2), 237-254.


\bibitem[Gross \& Reusken\ (2011)]{GrossReusken2011}
{\sc Gross, S. \& Reusken, A.} 2011 {Numerical Methods for Two-phase Incompressible Flows} {\bf Vol 1}, Springer, Berlin/Heidelberg.


\bibitem[Haines\ (1930)]{Haines1930}
{\sc Haines, W. B.} 1930 {Studies in the physical properties of soil. V. The hysteresis effect in capillary properties, and the modes of moisture distribution associated therewith}, {\it J. Agr. Sci.} {\bf 20}(1), 97-116.


\bibitem[Hu et al.\ (2019)]{HuEtAl2019}
{\sc Hu, R., Lan, T., Wei, G.-J. \& Chen, Y.-F.} 2019 {Phase diagram of quasi-static immiscible displacement in disordered porous media}, {\it J. Fluid Mech.} {\bf 875}, 448-475.


\bibitem[Jung et al.\ (2016)]{JungEtAl2016}
{\sc Jung, M., Brinkmann, M., Seemann, R., Hiller, T., Sancez de la Lama, M. \& Herminghaus, S.} 2016 {Wettability controls slow immiscible displacement through local interfacial instabilities}, {\it Phys. Rev. Fluids} {\bf 1}, 074202 1-19.


\bibitem[Kirchner et al.\ (2000)]{KirchnerEtAl2000}
{\sc Kirchner, J. W., Feng, X., \& Neal, C.} 2000 {Fractal stream chemistry and its implications for contaminant transport in catchments}, {\it Nature}, {\bf 403}(6769), 524-527.


\bibitem[Larson et al.\ (1977)]{LarsonEtAl1977}
{\sc Larson, R. G., Scriven, L. E. \& Davis, H. T.} 1977 {Percolation theory of residual phases in porous media}, {\it Nature} {\bf 268}, 409-413.


\bibitem[Linga\ (2025)]{Linga2025}
{\sc Linga, G.} Twoasis \texttt{https://github.com/gautelinga/twoasis}.


\bibitem[Linga et al.\ (2019)]{LingaEtAl2019}
{\sc Linga, G., Bolet, A., \& Mathiesen, J.} 2019 {Bernaise: A flexible framework for simulating two-phase electrohydrodynamic flows in complex domains}, {\it Front. Phys.}, {\bf 7}, 21.


\bibitem[Logg et al.\ (2012)]{LoggEtAl2012}
{\sc Logg, A., Mardal, K.-A. \& Wells, G.} 2012 {Automated solution of differential equations by the finite element method: The FEniCS book} {\bf Vol. 84}, Springer Science \& Business Media.


\bibitem[Mathiesen et al.\ (2023)]{MathiesenEtAl2023}
{\sc Mathiesen, J., Linga, G., Misztal, M., Renard, F. \& Le Borgne, T.} 2023 {Dynamic Fluid Connectivity Controls Solute Disperion in Multiphase Porous Media Flow}, {\it Geophys. Res. Lett.} \textbf{50}, e2023GL105233.


\bibitem[Melrose \& Brandner\ (1974)]{MelroseBrandner1974}
{\sc Melrose, J. C. \& Brandner, C. F.} 1974 {Role of Capillary Forces in Determining Microscopic Displacement Efficiency for Oil Recovery by Waterflooding}, {\it Can. J. Pet. Technol.} {\bf 13}, 54-62.


\bibitem[Mortensen \& Valen-Sendstad (2015)]{MortensenValenSendstad2015}
{\sc Mortensen, M., \& Valen-Sendstad, K.} 2015 {Oasis: a high-level/high-performance open source Navier–Stokes solver}, {\it Comp. Phys. Commun.}, {\bf 188}, 177-188.


\bibitem[Primkulov et al.\ (2021)]{PrimkulovEtAl2021}
{\sc Primkulov, B. K., Pahlavan, A. A., Fu, X., Zhao, B., MacMinn, C. W. \& Juanes, R.} 2021 {Wettability and LeNormand's diagram}, {\it J. Fluid Mech.} {\bf 923}, A34.


\bibitem[Rahbeh \& Mohtar\ (2007)]{RahbehMohtar2007}
{\sc Rahbeh, M. E., \& Mohtar, R. H.} 2007 {Application of multiphase transport models to field remediation by air sparging and soil vapor extraction}, {\it J. Haz. Mat.}, {\bf 143}(1-2), 156-170.


\bibitem[Shewchuk\ (1996)]{Shewchuk1996}
{\sc Shewchuk, J.} 1996 {Triangle: engineering a 2D quality mesh generator and delaunay triangulator}. In: Lin, M. C., Manocha, D. (editors) {\it Applied Computational Geometry: Towards Geometric Engineering. Lecture Notes in Computer Science} {\bf Vol. 1148}, 203-222. Springer, Berlin.


\bibitem[Singh et al.\ (2019)]{SinghEtAl2019}
{\sc Singh, K., Jung, M., Brinkmann, M. \& Seemann, R.} 2019 {Capillary-Dominated Fluid Displacement in Porous Media}, {\it Annu. Rev. Fluid Mech.} {\bf 51}, 429-449.


\bibitem[Taber\ (1969)]{Taber1969} 
{\sc Taber, J. J.} 1969 {Dynamic and Static Forces Required To Remove a Discontinuous Oil Phase from Porous Media Containing Both Oil and Water}, {\it Soc. Pet. Eng. J.} {\bf 9}, 3-12.  


\bibitem[Tallakstad et al.\ (2009)]{TallakstadEtAl2009}
{\sc Tallakstad, K. T., L{\o}voll, G., Knudsen, H. A., Ramstad, T., Flekk{\o}y, E. G. \& M{\aa}l{\o}y, K. J.} 2009 {Steady-state, simultaneous two-phase flow in porous media: An experimental study}, {\it Phys. Rev. E} {\bf 80}, 036308 1-13.


\bibitem[Wang et al.\ (2023)]{WangEtAl2023}
{\sc Wang, Z., Pereira, J.-M., Sauret, E. \& Gan, Y.} 2023, {Wettability impacts residual trapping of immiscible fluids during cycling injection}, {\it J. Fluid Mech.} {\bf 961}, A19.


\bibitem[Yue et al.\ (2007)]{YueEtAl2007}
{\sc Yue, P., Zhou, C., \& Feng, J. J.} 2007 {Spontaneous shrinkage of drops and mass conservation in phase-field simulations}, {\it J. Comput. Phys.}, {\bf 223}(1), 1-9.


\bibitem[Zhan \& Ng\ (2004)]{ZhanNg2004}
{\sc Zhan, T. L., \& Ng, C. W. } 2004 {Analytical analysis of rainfall infiltration mechanism in unsaturated soils}, {Int. J. Geomech.}, {\bf 4}(4), 273-284.





\end{thebibliography}
\end{document}